# Physics-Based Factorized Machine Learning for Predicting Ionic Dielectric Tensors

Atsushi Takigawa[1,2,4], Shin Kiyohara[1,4]* and Yu Kumagai[1,3]**

[1]Institute for Materials Research, Tohoku University, 2-1-1 Katahira, Aoba-ku, Sendai 980-8577, Japan

[2]Graduate School of Engineering, Tohoku University, 6-6 Aramaki Aoba, Aoba-ku, Sendai 980-8579, Japan

[3]Organization for Advanced Studies, Tohoku University, 2-1-1 Katahira, Aoba-ku, Sendai 980-8577, Japan

[4]These authors contributed equally: Atsushi Takigawa and Shin Kiyohara

* sin@tohoku.ac.jp

**yukumagai@tohoku.ac.jp

**ABSTRACT**. Considerable effort continues to be devoted to the exploration of next-generation high-κ materials that combine a high dielectric constant with a wide band gap. However, machine learning (ML)-based virtual screening has remained challenging, primarily due to the low accuracy in predicting the ionic contribution to the dielectric tensor, which dominates the dielectric performance of high-κ materials. We here propose a joint ML model that predicts Born effective charges using an equivariant graph neural network, and phonon properties using a highly accurate pretrained ML potential. The ionic dielectric tensor is then computed analytically from these quantities. This approach significantly improves the accuracy of ionic contribution. Using the proposed model, we successfully identified 31 novel high-κ oxides from a screening pool of over 8,000 candidates.

## I. INTRODUCTION

High-κ materials play an essential role in a wide range of advanced electronic and energy storage devices, e.g., transistors and capacitors [1–3]. In these applications, a high dielectric constant is essential for maximizing capacitance, while a wide band gap ($E_g$) remains necessary to ensure insulation and minimize leakage [3–6]. For example, in transistor gate insulators, $SiO_2$ has traditionally been the dominant material. Although its wide band gap ($E_g \approx 8.1$ eV) [7] is beneficial, the low dielectric constant ($\varepsilon \approx 3.8$) [7] requires thinning of the capacitor to realize high capacitance, which in turn leads to an increase in leakage current due to quantum tunneling [3]. In capacitor applications, perovskite oxides, such as $BaTiO_3$ and $SrTiO_3$, have been widely employed [8]. Yet, achieving both high permittivity and low leakage current in ultrathin perovskite oxide films remains a major challenge during device miniaturization [3,9]. In addition, they require high-temperature processing and are therefore difficult to integrate with CMOS technology [5,10].

In both applications, attention is now shifting to $HfO_2$ as it offers a relatively high dielectric constant ($\varepsilon \approx$ 20-30), a wide band gap ($E_g \approx 5.7$ eV), and excellent thermal and chemical stability [2,4,5,11–13]. However, $HfO_2$ exhibits a smaller band gap than $SiO_2$ and a lower conduction band offset with silicon, which can lead to increase leakage current [7,13]. Furthermore, its dielectric constant is lower than those of the other high-κ materials such as $BaTiO_3$, limiting the achievable capacitance. Thus, considerable effort has also been devoted to exploring next-generation high-κ materials that combine a high dielectric constant with a wide band gap [14,15], and this effort is still ongoing.

Discovering novel materials solely through experiments significantly constrains the candidate space. In recent years, high-throughput first-principles calculations have played an important role in guiding such explorations. Still, their high computational cost limits the scope of searches, and machine learning (ML)-based virtual screening has emerged as a promising alternative [16–18]. Although the band gap can now be predicted accurately by ML [19–21], the accuracy for dielectric constants remains relatively low, which lowers the success rate of virtual screening.

The static dielectric tensor ($\boldsymbol{\varepsilon}$) is described as the sum of the electronic ($\boldsymbol{\varepsilon}^{\mathrm{ele}}$) and the ionic ($\boldsymbol{\varepsilon}^{\mathrm{ion}}$) contributions, in addition to the vacuum permittivity: $\boldsymbol{\varepsilon} = \mathbf{1} + \boldsymbol{\varepsilon}^{\mathrm{ele}} + \boldsymbol{\varepsilon}^{\mathrm{ion}}$. Here, dielectric constants are described using the relative permittivity. The electronic part $\varepsilon_{\alpha\beta}^{\mathrm{ele}}$ is derived via the Kramers-Kronig relation:

$$\varepsilon_{\alpha\beta}^{\mathrm{ele}} = \frac{2}{\pi}\mathcal{P}\int_0^\infty \frac{\varepsilon_{\alpha\beta}^{\mathrm{img}}(\omega')}{\omega'}\,d\omega', \tag{1}$$

where $\alpha$ and $\beta$ denote directions in Cartesian coordinates, $\mathcal{P}$ denotes the principal value, and $\varepsilon_{\alpha\beta}^{\mathrm{img}}$ is the imaginary part of the dielectric function [22]. As seen in the integrand, the denominator $\omega'$ means that lower transition energy (i.e., smaller band gap) contributes more to $\varepsilon_{\alpha\beta}^{\mathrm{ele}}$ because $\varepsilon_{\alpha\beta}^{\mathrm{img}}$ vanishes below the band gap [23]. Therefore, $\boldsymbol{\varepsilon}^{\mathrm{ele}}$ exhibits a trade-off relationship with the band gap, as shown in Fig. 1(a).

In contrast, as illustrated in Fig. 1(b), the $\boldsymbol{\varepsilon}^{\mathrm{ion}}$ does not necessarily correlate with a band gap because it is determined by the Born effective charges (BECs) and phonon properties, as will be shown later. In addition, $\boldsymbol{\varepsilon}^{\mathrm{ion}}$ can reach significantly higher values, such as several hundred, meaning that accurately predicting $\boldsymbol{\varepsilon}^{\mathrm{ion}}$ is pivotal for screening materials that exhibit a high dielectric constant.

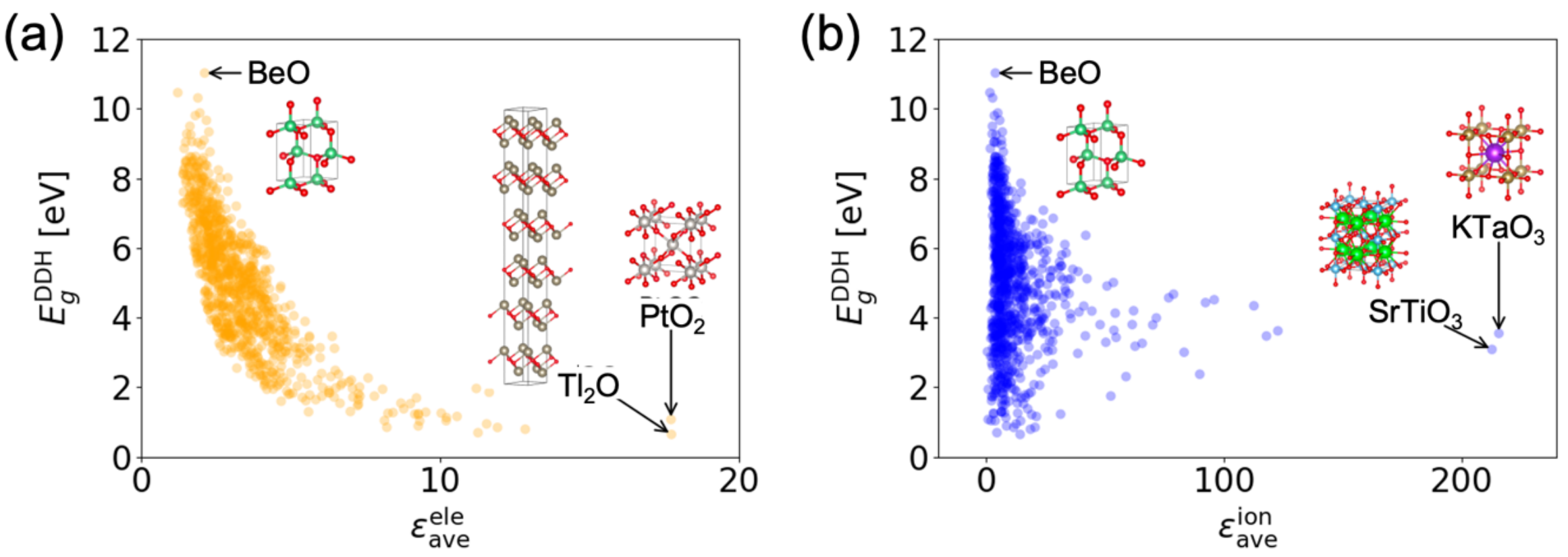

FIG. 1. Relationship between dielectric constants and band gaps. Spherically averaged (a) electronic ($\varepsilon_{\mathrm{ave}}^{\mathrm{ele}}$) and (b) ionic ($\varepsilon_{\mathrm{ave}}^{\mathrm{ion}}$) dielectric constants are plotted against band gaps ($E_{\mathrm{g}}^{\mathrm{DDH}}$). The dielectric constants were computed using PBEsol, while the band gaps were obtained using the DDH (see text for details). All data are derived from first-principles calculations performed on the 928 oxides.

Several studies have reported ML-based predictions of $\boldsymbol{\varepsilon}^{\mathrm{ele}}$ and $\boldsymbol{\varepsilon}^{\mathrm{ion}}$. Despite the diversity in target materials (e.g., oxides [23–25] and other materials [18,26,27]), dielectric properties (scalars [23–28] or tensors [18]), and ML models (including random forest [23,25], support vector regression [24,27], gradient-boosting-based regression [26], and graph neural networks (GNN) [18,27–29]), the prediction accuracy for $\boldsymbol{\varepsilon}^{\text{ion}}$ consistently lags behind that of $\boldsymbol{\varepsilon}^{\text{ele}}$. In particular, the predictions tend to be less accurate for materials with higher $\boldsymbol{\varepsilon}^{\text{ion}}$ as they are likely to lie in the extrapolation regime. Thus, the previous prediction models are hardly applicable to the screening of high-κ materials.

In this work, we develop a high-accuracy ML model for predicting $\boldsymbol{\varepsilon}^{\text{ion}}$ by employing separate ML models for BEC tensors and phonon properties and integrating these predicted quantities to compute $\boldsymbol{\varepsilon}^{\text{ion}}$ analytically. We focus on oxides since they are usually easy to synthesize under ambient conditions and have larger band gaps. The commercially used and intensively investigated dielectrics are indeed mainly oxides such as $SiO_2$, $BaTiO_3$, and $HfO_2$. Leveraging our framework, we conducted a virtual screening of 8,717 oxides and identified 31 high-κ oxides.

## II. RESULTS

### A. ML predictions of ionic dielectric tensors

The ionic dielectric tensor $\varepsilon_{\alpha\beta}^{\text{ion}}$ is given by the following expression [30,31]:

$$\varepsilon_{\alpha\beta}^{ion} = \frac{4\pi}{V_0} \sum_{j\alpha',j'\beta'} Z_{j,\alpha\alpha'}^{*} (\boldsymbol{\Phi}^{-1})_{j\alpha',j'\beta'} Z_{j',\beta\beta'}^{*}, \tag{2}$$

where $V_0$ is the volume of the unit cell, $\alpha$, $\alpha'$, $\beta$ and $\beta'$ are directions in Cartesian coordinates, $j$ is the atom index within the unit cell. $Z_{j,\alpha\alpha'}^{*}$ denotes the BEC tensor for ion $j$, and $(\boldsymbol{\Phi}^{-1})_{j\alpha',j'\beta'}$ is the inverse of the interatomic force constant matrix between ion $j$ and $j'$, respectively (see *Methods*). The BEC tensor is defined as

$$Z_{j,\alpha\beta}^{*} = \frac{V_0}{e} \frac{\partial P_\alpha}{\partial u_{j\beta}}, \tag{3}$$

where $e$ is the elementary charge, $P_\alpha$ is the polarization along direction $\alpha$, and $u_{j\beta}$ is the displacement of atom $j$ along $\beta$. It should be noted that the BEC tensors, being 3×3 second-rank tensors, are in general *non-symmetric*, as they are defined as second derivatives of the total energy with respect to atomic displacements and applied electric fields [30]. In first-principles calculations, they are calculated based on the density functional perturbation theory (DFPT). The limited accuracy of previous ML models for the $\boldsymbol{\varepsilon}^{\text{ion}}$ can be attributed to the inherent difficulty of simultaneously capturing both BECs and phonon properties within a single ML model. To overcome this, we predict each separately and analytically combine their outputs using Eq. (2) as shown in Fig. 2(a). We refer to this approach as the *joint* model.

High-κ materials often exhibit anomalously large BECs. For instance, in cubic $SrTiO_3$, the BECs of Ti and O along the Ti-O bonding direction are +7.26 and -5.73, respectively, which significantly deviate from their nominal ionic charges of +4 and -2 [32]. Such large deviation is known to be related to the dynamic charges [30,33]. In Supplemental Fig. 1 [34], we show distributions of the BECs for each element in our dataset. In an extreme case, the BEC for an oxygen atom in $NbVO_5$ reaches as high as -8.80. Thus, they must be accurately inferred via ML from the surrounding atomic environments.

We developed an ML model for the BECs that is applicable to a broad spectrum of oxides based on an E(3)-Equivariant GNN (EGNN) [35–37] (see *Methods* and Appendix A). The model was trained on first-principles–calculated data. The dataset, containing 928 oxides composed of 45 elements, was retrieved from our previous studies [38], and covers wide range of crystal structures (see Supplemental Figs. 2-3 [34]). The EGNN performance was evaluated using 5-fold cross-validation. The prediction accuracy was assessed by comparing the eigenvalues of the BEC tensors. Figure 2(b) presents the performance of our EGNN model, showing an excellent agreement with an average coefficient of determination ($R^2$) of 0.982 and root mean squared error (RMSE) of 0.373. While previous approaches for predicting the BECs are typically restricted to a single material [39–41] or a limited class of materials such as perovskites [42], our work is the first to demonstrate that BECs can be predicted with consistently high accuracy across a broad range of chemical compositions and crystal structures (see Appendix B). Note that the acoustic sum rule [30,43] is not explicitly enforced in the EGNN for the BECs; nevertheless, the predicted BECs satisfy this rule to high accuracy owing to the overall predictive fidelity of the model (see Appendix C).

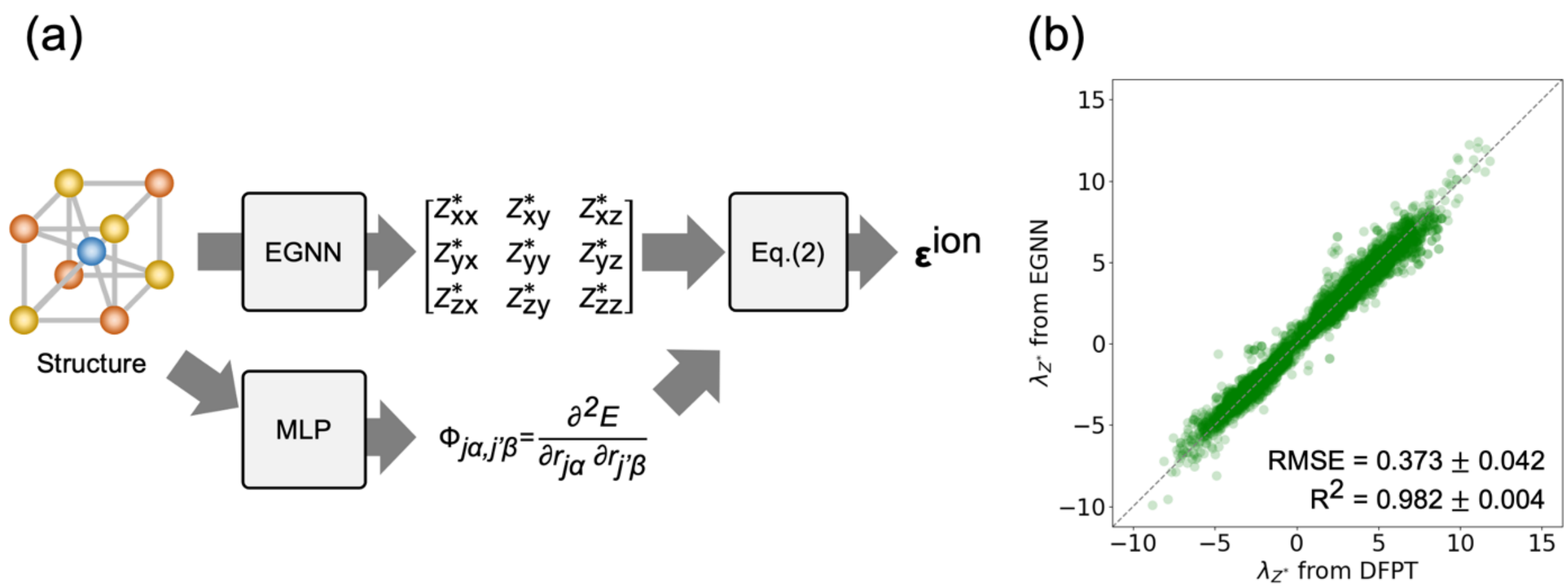

FIG. 2. (a) Schematic workflow for predicting the $\boldsymbol{\varepsilon}^{ion}$ in our joint model. From a given crystal structure, an EGNN is used to predict the BEC tensors ($\mathbf{Z}^*$) at all the atomic sites, while an MLP provides the force constant matrix ($\boldsymbol{\Phi}$). These quantities are then substituted into Eq. (2) to compute the $\boldsymbol{\varepsilon}^{ion}$. (b) Parity plot of the eigenvalues of $\mathbf{Z}^*$ ($\lambda_{\mathbf{Z}^*}$) between the reference DFPT values and ML predicted values (14,202 inequivalent data points at 4734 sites in the 928 oxides). Results from all test sets in 5-fold cross-validation are shown.

Another requirement to predict $\boldsymbol{\varepsilon}^{ion}$ via Eq. (2) is an accurate prediction of phonon properties. Recent advances in machine learning potentials (MLPs) have substantially enhanced the accuracy of their predictions in a wide range of material systems [44–49]. However, the evaluation of $\boldsymbol{\varepsilon}^{ion}$ is particularly sensitive to the accuracy of low-frequency phonon modes. When the frequencies of optical phonon modes become marginally unstable, the $\boldsymbol{\varepsilon}^{ion}$ diverges, as seen in Eq. (7) in the *Methods* section. To assess the performance of MLPs in predicting phonon properties, we evaluated four pretrained MLPs with a particular focus on their ability to predict the $\boldsymbol{\varepsilon}^{ion}$. To isolate the error originating from the MLPs, we calculated the $\boldsymbol{\varepsilon}^{ion}$ using the force constants predicted by the MLPs together with the BECs obtained from first-principles calculations.

The MLPs include eSEN-uma-s-1-omat (eSEN) [44], ORB-v3 (ORB) [45], SevenNet-MF-ompa (SevenNet) [47,49], and MACE-MPA-0 (MACE) [46]. Based on the Matbench discovery [50], the symmetric relative mean errors in predicted phonon mode contributions to thermal conductivity of eSEN, ORB, SevenNet, and MACE are 0.170, 0.210, 0.317, and 0.412, respectively.

Initially, we discuss the dynamical stability. In this study, a structure is considered dynamically unstable when the fourth lowest phonon frequency, $f_{4\mathrm{th}}$, at the Γpoint is smaller than 0.2 THz. Note that the training datasets of these MLPs were generated using the PBE functional [51], while we adopted the PBEsol functional [52]. Partly due to these, some structures were predicted to be unstable at the Γ point although they are stable in our PBEsol-based DFPT calculations. The numbers of such dynamically unstable oxides are 87, 62, 88 and 89 when using eSEN, ORB, SevenNet, and MACE, respectively. In principle, their $\boldsymbol{\varepsilon}^{\mathrm{ion}}$ values are not calculated because imaginary modes at the Γ point give unphysical negative contribution to the $\boldsymbol{\varepsilon}^{\mathrm{ion}}$ as shown in Eq. (7). Thus, we evaluated the predictive accuracy using 786 oxides, which are predicted to be dynamically stable by all the MLPs. Consequently, we realized that SevenNet achieves the highest accuracy for predicting $\boldsymbol{\varepsilon}^{\mathrm{ion}}$, although the differences among the MLPs are marginal as shown in Table 1 (see Supplemental Fig. 4 [34] for more details). Accordingly, SevenNet was selected for subsequent calculations involving phonon properties.

TABLE 1. $R^2$ and RMSE metrics for the contribution of the predicted force constant matrices to $\boldsymbol{\varepsilon}^{\mathrm{ion}}$. The evaluation was performed on 786 oxides in the training dataset, which were predicted by all MLPs to be dynamically stable at the Γ point. $\boldsymbol{\varepsilon}^{\mathrm{ion}}$ were estimated based on the force constant matrices predicted by MLPs, and the BECs obtained from DFPT. The predicted $\boldsymbol{\varepsilon}^{\mathrm{ion}}$ were converted to eigenvalues, followed by a base-10 logarithmic conversion, and subsequently compared with the corresponding reference DFPT values. The MLP achieving the highest accuracy is highlighted in bold.

| MLP model | $R^2$ | RMSE |
| --- | --- | --- |
| eSEN-uma-s-1-omat | 0.965 | 0.062 |
| ORB-v3 | 0.960 | 0.066 |
| MACE-MPA-0 | 0.952 | 0.073 |
| **SevenNet-MF-ompa** | **0.968** | **0.059** |

Finally, we predicted the $\boldsymbol{\varepsilon}^{\mathrm{ion}}$ via Eq. (2) using the EGNN for the BECs and the SevenNet for phonon properties (the joint model). Its predictive performance, evaluated using 5-fold cross-validation, was compared with that of a model that directly predicts the $\boldsymbol{\varepsilon}^{\mathrm{ion}}$ (referred to as the direct model), as shown in Fig. 3 (see Supplemental Note 1 [34] for more details). The accuracy metrics were evaluated using 839 oxides that were predicted to be dynamically stable at the Γ point by SevenNet. The direct model exhibits limited predictive capability, yielding an average $R^2$ of 0.411 and an average RMSE of 0.254. In contrast, the joint model achieves substantially improved accuracy, achieving a higher $R^2$ of 0.911 and a lower RMSE of 0.099. Although the direct comparison with previous studies [18,23–28] is not strictly appropriate due to differences in datasets and the target material systems, our joint model demonstrates the highest performance in predicting the $\boldsymbol{\varepsilon}^{\mathrm{ion}}$, highlighting the effectiveness of our joint model (see Appendix D).

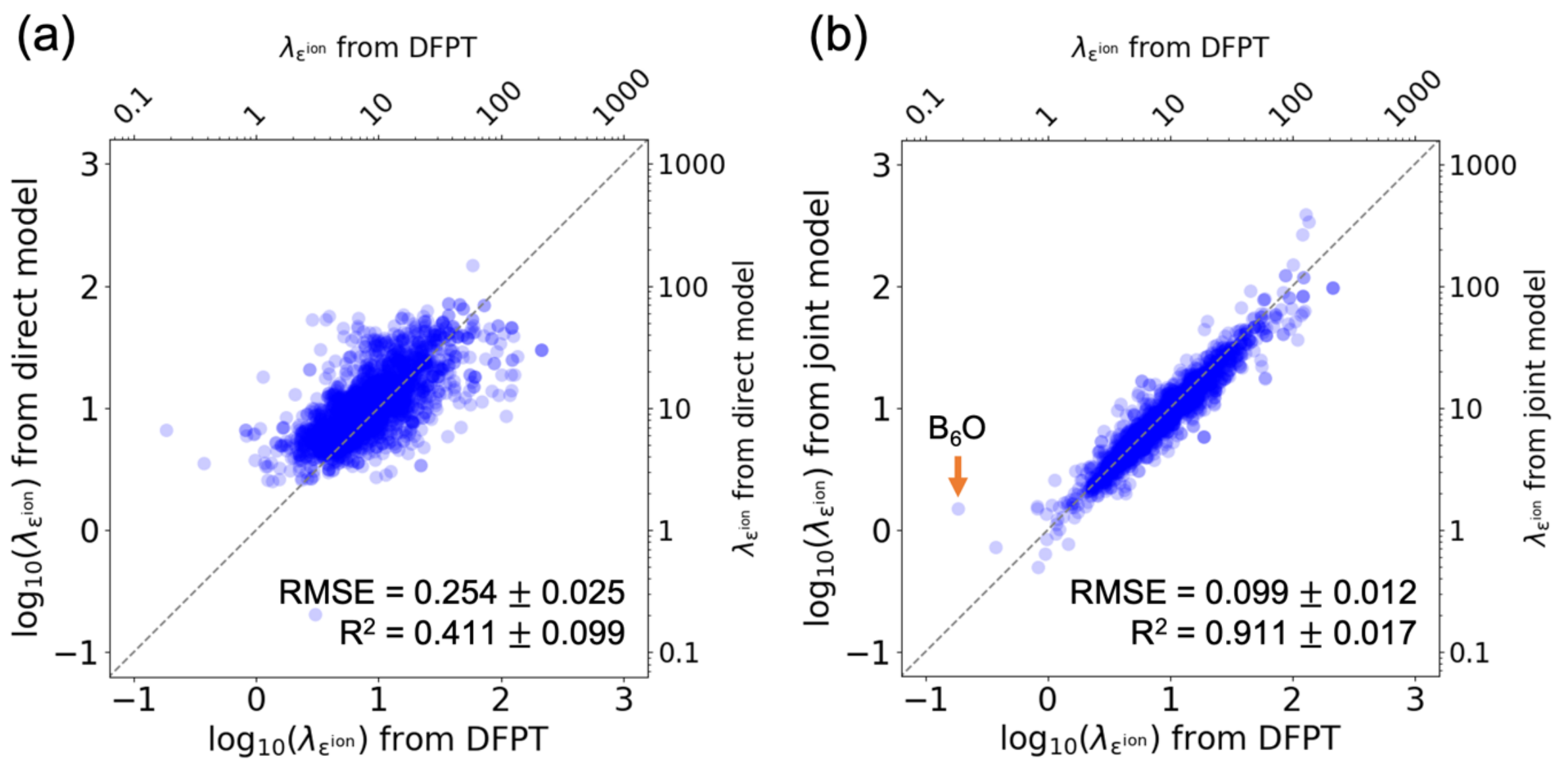

FIG. 3. Performance comparison of the conventional direct model and the proposed joint model. Parity plots of the eigenvalues of the ionic dielectric tensors are shown, comparing the reference DFPT values with the ML-predicted values obtained using (a) the direct model and (b) the joint model. The prediction accuracy was evaluated for the 839 oxides, which were predicted to be dynamically stable by SevenNet. For comparison, the ionic dielectric tensors were converted to eigenvalues ($\lambda_{\varepsilon^{\mathrm{ion}}}$) and transformed using a base-10 logarithm to mitigate scale differences. One predicted value of $\lambda_{\varepsilon^{\mathrm{ion}}}$ for $Ba_3Nb_2ZnO_9$, specifically -5.135, is omitted from panel (a) due to its physically ill-defined nature.

To closely examine the sources of the prediction errors, we decomposed them into two contributions: (1) combinations of ML-predicted BECs with DFPT-calculated phonons, and (2) combinations of DFPT-calculated BECs with ML-predicted phonons. This decomposition allows us to isolate how inaccuracies in each component propagate into the final prediction of $\boldsymbol{\varepsilon}^{\mathrm{ion}}$. The results were further classified into low- and high-$\boldsymbol{\varepsilon}^{\mathrm{ion}}$ regimes using $\lambda_{\boldsymbol{\varepsilon}^{\mathrm{ion}}} = 10$ as the dividing threshold, where $\lambda_{\boldsymbol{\varepsilon}^{\mathrm{ion}}}$ denotes an eigenvalue of the ionic dielectric tensor. The resulting BEC- and phonon-related contributions to the prediction errors are summarized in Table 2. We emphasize that the BEC- and phonon-related errors do not necessarily sum to the total prediction error, as partial cancellation between the two contributions can occur.

While the overall errors arise from both BECs and phonon contributions, in the high-$\boldsymbol{\varepsilon}^{\mathrm{ion}}$ regime, the prediction errors are predominantly governed by phonon-related inaccuracies. This is consistent with the inverse relationship between the ionic dielectric constants and phonon frequencies, as shown in Eq. (7). A single outlier appears in the low-$\boldsymbol{\varepsilon}^{\mathrm{ion}}$ regime: cubic $B_6O$, which exhibits an exceptionally small $\lambda_{\boldsymbol{\varepsilon}^{\mathrm{ion}}}$ of 0.3. Accurate prediction of such small $\boldsymbol{\varepsilon}^{\mathrm{ion}}$ values require highly precise predictions of the BECs. Indeed, the ML-predicted BECs show relatively large deviations from the DFPT values, which arises not only from the intrinsically small magnitude of BECs but also from the lack of chemically similar environments in the training dataset. Further details of the error decomposition are provided in Supplemental Note 2 [34].

TABLE 2. Decomposition of $\boldsymbol{\varepsilon}^{\text{ion}}$ prediction errors by SevenNet into contributions from the BECs and phonon properties. RMSEs were computed for the base-10 logarithm of the eigenvalues of $\boldsymbol{\varepsilon}^{\text{ion}}$ ($\lambda_{\boldsymbol{\varepsilon}^{\text{ion}}}$). Results are reported for a dataset comprising 839 dynamically stable oxides. The corresponding mean absolute error (MAE) and $R^2$ are provided in Supplemental Note 2 [34]. RMSEs in the high- and low-$\boldsymbol{\varepsilon}^{\text{ion}}$ regions, separated by the threshold $\lambda_{\boldsymbol{\varepsilon}^{\text{ion}}} = 10$, are also tabulated. The "Total" column reports the prediction errors of the joint model, whereas the "Phonon" and "BEC" columns represent the phonon- and BEC-related errors, respectively (see text for details).

| Dataset | Total | Phonon | BEC |
|---|---|---|---|
| All | 0.100 | 0.064 | 0.071 |
| High-$\boldsymbol{\varepsilon}^{\text{ion}}$ | 0.120 | 0.092 | 0.075 |
| Low-$\boldsymbol{\varepsilon}^{\text{ion}}$ | 0.086 | 0.041 | 0.069 |

### B. Virtual screening of high-κ oxides

Using our joint model, we performed a virtual screening of promising high-κ oxides from a candidate pool extracted from the Materials Project database (MPD) [53]. A stepwise screening procedure was employed, as illustrated in Fig. 4(a). Specifically, we considered oxides for which dielectric constants had not yet been reported but for which band gaps were available. The candidate pool was further restricted to compositions containing only the 45 elements covered by our EGNN for prediction of BECs, and the 928 oxides used in the training dataset were excluded to avoid the data leakage. Application of these criteria yielded 19,569 oxides from the MPD.

For virtual screening, target oxides were retrieved from the MPD on May 20, 2025; these oxides are composed of elements that appear in the oxides of the training dataset. To avoid the data leakage, the 928 oxides used in the training dataset were excluded, resulting in 19,569 oxides.

We then removed metallic oxides, reducing the number of candidates from 19,569 to 13,261. Magnetic oxides were further excluded, as dielectric tensors computed for spin-polarized systems can be highly sensitive to the choice of effective $U$ parameters and to specific spin configurations, thereby limiting the predictive reliability for such magnetic materials. As a result, the final candidate set comprised 8,717 oxides. The distributions of elements in the training dataset and the screening candidates exhibit substantial overlap, supporting the applicability of our joint model to virtual screening. (see Supplemental Figs. 2-3 [34]).

We firstly predicted the phonon properties at the Γ point using the SevenNet and excluded dynamically unstable oxides. Consequently, 4,654 oxides survived.

In the second stage, we performed screening based on the r$^2$SCAN band gaps reported in the MPD and the ML-predicted dielectric constants. Although the r$^2$SCAN band gaps underestimate experimental values by approximately 50% (see Supplemental Fig. 5 [34]), their strong correlation with experimental values supports their use for screening purposes. To evaluate the static dielectric tensor ($\boldsymbol{\varepsilon}$), both $\boldsymbol{\varepsilon}^{\text{ele}}$ and $\boldsymbol{\varepsilon}^{\text{ion}}$ are required. Accordingly, we additionally constructed an EGNN model to predict the $\boldsymbol{\varepsilon}^{\text{ele}}$. The prediction accuracy was assessed using 5-fold cross-validation for the 839 oxides predicted to be dynamically stable by SevenNet. For $\log_{10}(\lambda_{\boldsymbol{\varepsilon}^{\text{ele}}})$ and $\log_{10}(\lambda_{\boldsymbol{\varepsilon}})$, the models achieved an average $R^2$ values of 0.714 and 0.905, respectively (see *Methods* and Supplemental Note 1 [34]). Finally, the EGNNs for both the $\boldsymbol{\varepsilon}^{\text{ele}}$ and BECs were retrained using full dataset of 928 oxides to further improve the predictive accuracy.

Since superior dielectric materials exhibit both a high dielectric constant and a wide band gap, we introduced a figure of merit (FoM) that captures these two key properties simultaneously. Here, FoM was defined as $\varepsilon_{\text{ave}} \cdot E_{\text{g}}$, where $\varepsilon_{\text{ave}}$ denotes spherically averaged static dielectric constant, and the band gap $E_{\text{g}}$ is in eV, following previous studies [14,15,17,18]. Using these ML-predicted dielectric constants ($\varepsilon_{\text{ave}}^{\text{ML}}$) and r²SCAN band gaps ($E_{\text{g}}^{\text{r}^2\text{SCAN}}$), we calculated the FoM, which is referred to as $\text{FoM}_{\text{r}^2\text{SCAN}}^{\text{ML}}$, for each oxide and applied a screening threshold of $\text{FoM}_{\text{r}^2\text{SCAN}}^{\text{ML}} \geq 200$. Consequently, 142 oxides remained.

In the third stage, we conducted DFPT calculations using PBEsol to validate both the predicted $\varepsilon_{\text{ave}}^{\text{ML}}$ and the dynamical stability of the screened candidates. Candidates for which the first-principles calculations failed to converge were also excluded. This step reduced the pool to 121 oxides. We then calculated the band gaps using dielectric-dependent hybrid functional (DDH) calculations [54], which predict the band gaps of oxides accurately (see Supplemental Fig. 5 [34]). Applying a more stringent $\text{FoM}_{\text{DDH}}^{\text{DFPT}} \geq 350$ criterion subsequently reduced the number of candidates to 31. Identified novel high-κ oxides are plotted together with the training data in Fig. 4(b).

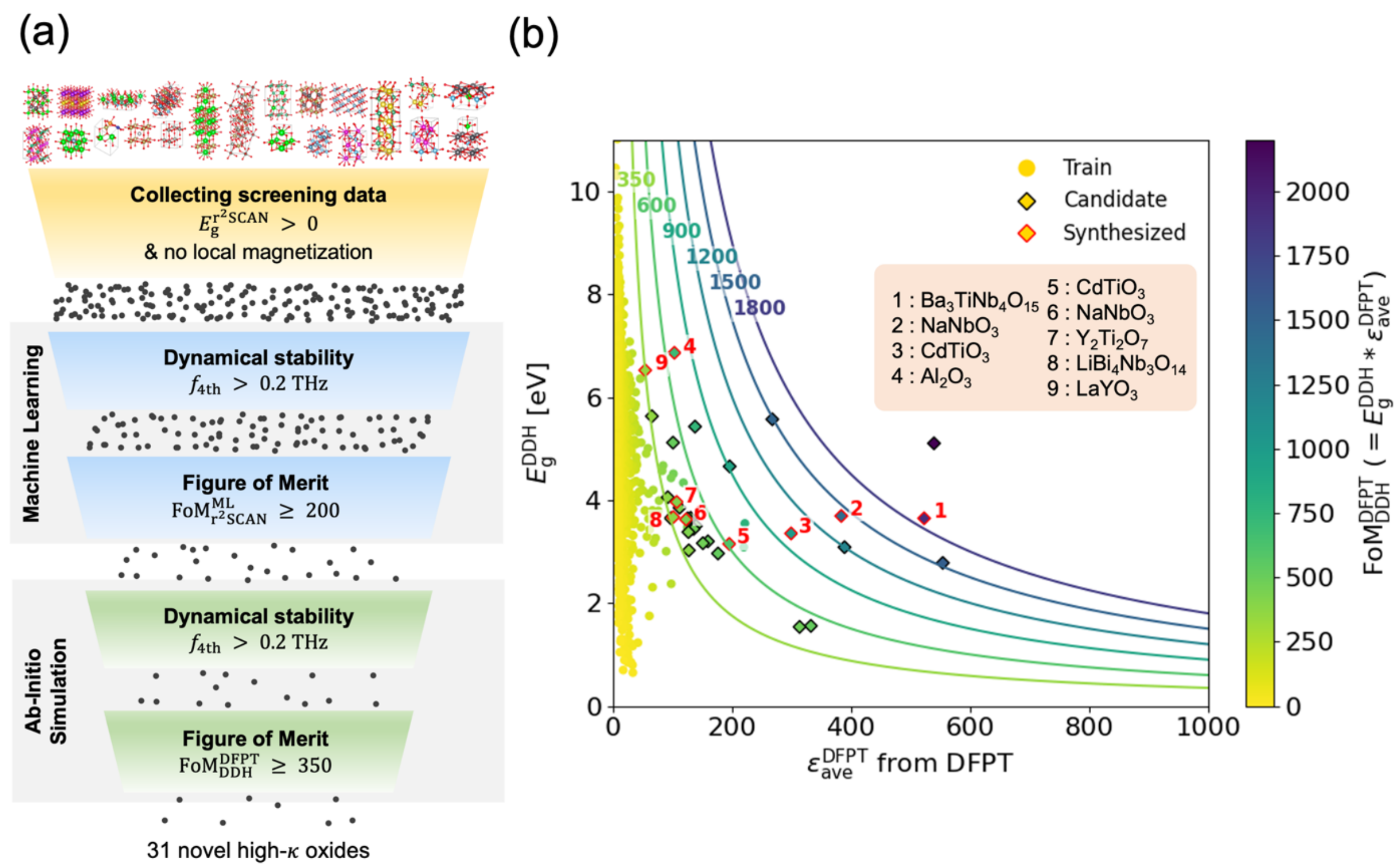

FIG. 4. (a) Schematic of the screening workflow to identify novel high-κ oxides. Details are provided in the main text and the *Methods* section. (b) Plot of the spherically averaged dielectric constant ($\varepsilon_{\text{ave}}^{\text{DFPT}}$) versus the band gap ($E_{\text{g}}^{\text{DDH}}$) for the 928 oxides in the ML training dataset and the 31 high-κ oxides identified by virtual screening ($\text{FoM}_{\text{DDH}}^{\text{DFPT}} \geq 350$). The dielectric constants were calculated using PBEsol-based DFPT, and the band gaps were computed using the DDH.

The color scale represents the $\mathrm{FoM}_{\mathrm{DDH}}^{\mathrm{DFPT}}$. Circles denote oxides in the training set, while diamonds indicate the identified high-κ oxides in this study. Among the diamonds, those enclosed in red lines correspond to synthesized oxides, whose formula are also described in the inset.

Among the training data, 9 out of 928 oxides exhibited $\mathrm{FoM}_{\mathrm{DDH}}^{\mathrm{DFPT}} \geq 350$ (see Table 3). In contrast, among 142 high-κ oxides with $\mathrm{FoM}_{\mathrm{r^2SCAN}}^{\mathrm{ML}} \geq 200$, 31 oxides satisfy $\mathrm{FoM}_{\mathrm{DDH}}^{\mathrm{DFPT}} \geq 350$. The properties of these 31 candidates are summarized in Table 3, which lists the top ten oxides ranked by $\mathrm{FoM}_{\mathrm{DDH}}^{\mathrm{DFPT}}$, together with those that are either experimentally synthesized or exhibit an energy above the convex hull, $E_{\mathrm{hull}}$ (in eV/atom), of zero. A complete list of the 31 identified high-κ oxides, along with the nine oxides from the training dataset, is provided in Supplemental Note 3 [34]. An important benchmark is $KTaO_3$, which exhibits the highest $\mathrm{FoM}_{\mathrm{DDH}}^{\mathrm{DFPT}}$ among the 928 training oxides ($E_{\mathrm{g}}^{\mathrm{DDH}}$ = 3.5 eV, $\varepsilon_{\mathrm{ave}}^{\mathrm{DFPT}}$ = 221, and $\mathrm{FoM}_{\mathrm{DDH}}^{\mathrm{DFPT}}$ = 785). Among the 31 identified oxides, eight exhibit higher $\mathrm{FoM}_{\mathrm{DDH}}^{\mathrm{DFPT}}$ values than $KTaO_3$, demonstrating a marked improvement over the best-performing oxide in the training dataset.

TABLE 3. Properties of identified novel high-κ oxides as well as those in the training dataset, including the calculated band gaps ($E_{\mathrm{g}}^{\mathrm{DDH}}$, in eV), spherically averaged electronic and ionic dielectric constants ($\varepsilon_{\mathrm{ave}}^{\mathrm{ele,DFPT}}$, $\varepsilon_{\mathrm{ave}}^{\mathrm{ion,DFPT}}$), and figures of merit ($\mathrm{FoM}_{\mathrm{DDH}}^{\mathrm{DFPT}}$). Band gaps were computed using the DDH, whereas dielectric constants were calculated using PBEsol. The "Group" column classifies each oxide as follows: "Top" denotes the top 10 screened oxides ranked by $\mathrm{FoM}_{\mathrm{DDH}}^{\mathrm{DFPT}}$; "Exp" corresponds to oxides that have been experimentally reported; "Stable" exhibits the oxides with an energy above the convex hull ($E_{\mathrm{hull}}$) of zero; "Train" indicates oxides from the training dataset. The "$E_{\mathrm{hull}}$" column corresponds to the values reported in the MPD, and "--" means that $E_{\mathrm{hull}}$ is zero. The column "Synthesized" denotes whether the oxide has been experimentally synthesized. When available, experimental dielectric constants ($\varepsilon_{\mathrm{ave}}^{\mathrm{Exp}}$) and characteristic crystal structure names (e.g., perovskite, pyrochlore) are noted. References for previously reported syntheses and experimental dielectric constants are also provided in Supplemental Note 3 [34].

| Group | Formula | $E_g^{DDH}$ | $\varepsilon_{ave}^{ele,DFPT}$ | $\varepsilon_{ave}^{ion,DFPT}$ | $FoM_{DDH}^{DFPT}$ | $E_{hull}$ | Synthesized | $\varepsilon_{ave}^{Exp}$ | Structure |
|---|---|---|---|---|---|---|---|---|---|
| Top | $SrZrO_3$ | 5.1 | 3.7 | 535 | 2752 | 0.02 | | | Perovskite |
| | $Ba_3TiNb_4O_{15}$ | 3.7 | 5.0 | 516 | 1912 | 0.02 | ✓ [55] | 900 [56–60] | $ReO_3$ |
| | $Pb_2TiZrO_6$ | 2.8 | 5.8 | 547 | 1540 | 0.04 | | | Perovskite |
| | $Sr_3Hf_2O_7$ | 5.6 | 3.2 | 263 | 1490 | 0.02 | | | Layered perovskite |
| | $NaNbO_3$ | 3.7 | 4.8 | 377 | 1421 | 0.00 | ✓ [61–64] | 200 [65] | Perovskite |
| | $AgTaO_3$ | 3.1 | 5.7 | 382 | 1200 | 0.05 | | | Perovskite |
| | $CdTiO_3$ | 3.4 | 6.7 | 290 | 1001 | 0.04 | ✓ [66] | | Perovskite |
| | $Ta_2Zn_4O_9$ | 4.7 | 4.2 | 190 | 911 | 0.10 | | | |
| | $Sr_3CaO_4$ | 5.4 | 2.9 | 134 | 747 | 0.03 | | | Caswellsilverite-like |
| | $Al_2O_3$ | 6.9 | 2.5 | 98 | 699 | 0.28 | ✓ [67] | | $CaIrO_3$ |
| Exp | $CdTiO_3$ | 3.2 | 6.7 | 186 | 611 | 0.04 | ✓ [68–70] | 2100 [71] | Perovskite |
| | $NaNbO_3$ | 3.6 | 4.9 | 115 | 439 | 0.00 | ✓ [61] | | Perovskite |
| | $Y_2Ti_2O_7$ | 4.0 | 5.2 | 99 | 416 | 0.01 | ✓ [72–78] | 54 [79] | Pyrochlore |
| | $LiBi_4Nb_3O_{14}$ | 3.7 | 5.8 | 94 | 371 | -- | ✓ [80] | | Aurivillius-derived |
| | $LaYO_3$ | 6.5 | 3.6 | 49 | 351 | 0.04 | ✓ [81] | | Perovskite |
| Stable | $Ta_4Cd_3HgO_{14}$ | 3.0 | 5.0 | 170 | 522 | -- | | | |
| | $SrTa_4Cd_3O_{14}$ | 3.9 | 4.5 | 106 | 430 | -- | | | |
| | $Ta_2CdPbO_7$ | 3.4 | 5.2 | 120 | 427 | -- | | | |
| | $Ta_2HgPbO_7$ | 3.0 | 5.5 | 121 | 384 | -- | | | |
| | $CaTa_4Cd_3O_{14}$ | 4.1 | 4.5 | 86 | 372 | -- | | | |
| Train | $KTaO_3$ | 3.5 | 4.5 | 216 | 785 | -- | ✓ [82,83] | 5000 [84] | Perovskite |
| | $SrTiO_3$ | 3.1 | 5.7 | 213 | 678 | -- | ✓ [85–97] | 4500 [98] | Perovskite |
| | $NaTaO_3$ | 4.3 | 4.3 | 113 | 512 | -- | ✓ [99–102] | 147 [103] | Perovskite |
| | $Cd_2Ta_2O_7$ | 3.6 | 4.8 | 123 | 466 | -- | ✓ [104,105] | 1200 [106] | Perovskite |
| | $Ba_3Nb_2CdO_9$ | 4.5 | 4.2 | 96 | 457 | -- | | | Perovskite-derived |
| | $CaTiO_3$ | 3.5 | 5.6 | 118 | 433 | -- | ✓ [69,107–123] | 180 [124] | Perovskite |
| | $Ba_3CaNb_2O_9$ | 4.4 | 4.0 | 92 | 431 | -- | ✓ [125] | 40 [126] | Perovskite-derived |
| | $KNbWO_6$ | 4.7 | 4.0 | 79 | 392 | -- | ✓ [127] | | Pyrochlore-derived |
| | $TaTlWO_6$ | 4.6 | 4.4 | 77 | 376 | -- | | | |

For both the nine oxides in the training dataset and the 31 identified high-κ oxides with $\mathrm{FoM}_{\mathrm{DDH}}^{\mathrm{DFPT}} \geq 350$, the ratios of ionic contributions to the total dielectric constants exceed 90%, reinforcing the notion that accurate prediction of $\boldsymbol{\varepsilon}^{\mathrm{ion}}$ is the key to identifying novel high-κ oxides.

Of the 31 identified oxides, experimental synthesis has already been reported for 9 oxides. For four of these oxides, experimental values of dielectric constant ($\varepsilon_{\mathrm{ave}}^{\mathrm{Exp}}$) have been reported, and all four exhibit high $\varepsilon_{\mathrm{ave}}^{\mathrm{Exp}}$ values, consistent with our predictions. For the remaining five oxides, no $\varepsilon_{\mathrm{ave}}^{\mathrm{Exp}}$ have been reported. Although $CdTiO_3$ in the *Pmc2₁* phase has been experimentally synthesized [66], previous studies suggest that alternative phases, *Pbnm* at room temperature and *Pna2₁* below 80 K, are more stable [68–70]. $NaNbO_3$ in the *Pna2₁* phase was investigated by Parker *et al.*, who identified the *Pna2₁* space group as the most suitable structure for its polar phase [61–64]. $Al_2O_3$ in the *Cmcm* phase adopts the $CaIrO_3$-type post-perovskite structure, which typically forms only under high-pressure conditions [67]. In contrast, $LiBi_4Nb_3O_{14}$ and $LaYO_3$ have been successfully synthesized under ambient conditions [80,81]. $LiBi_4Nb_3O_{14}$ has been studies as a photocatalyst, though its dielectric properties have not yet been measured. $LaYO_3$ possesses a perovskite structure and has been predicted as a promising high-κ oxide by Coh et al. [128], but no experimental $\varepsilon_{\mathrm{ave}}$ has been reported to date. Thus, $LiBi_4Nb_3O_{14}$ and $LaYO_3$ are experimentally accessible as-yet-unexplored high-κ oxides that require future experimental validation.

Analysis of the 31 identified oxides revealed that 29 include transition metals with empty *d*-orbitals, namely $Y^{3+}$, $Ti^{4+}$, $Zr^{4+}$, $Hf^{4+}$, $Nb^{5+}$, $Ta^{5+}$, and $W^{6+}$. This makes sense as such $d^0$-ness condition is commonly associated with large dynamic charges and, consequently, an enhanced dielectric response [129]. Given that such $d^0$ oxides exhibit anomalous BECs that deviate significantly from the nominal ionic charges, our EGNN predictions are indispensable for reliably identifying these oxides. Exceptions are only $CaIrO_3$-type $Al_2O_3$, and $Sr_3CaO_4$. Unfortunately, their syntheses under ambient conditions are expected to be difficult, as the former case is a high-pressure phase [67], while the latter is a theoretically predicted structure [130] and is expected to show strong cation disordering.

Almost all final candidates are composed of oxygen octahedra network centered around transition metal cations. Such octahedral frameworks are commonly observed in crystal structures that exhibit high dielectric constants, including perovskite structures, such as $BaTiO_3$ and $SrTiO_3$ [8]. The perovskite-related frameworks identified in this study encompass distorted perovskites, layered perovskites, post-perovskites, pyrochlores, and $ReO_3$ structures. $LiBi_4Nb_3O_{14}$ is specifically classified as a variant of the Aurivillius structure, which is a layered perovskite. The predicted high dielectric constants are primarily attributed to their structural flexibility, particularly the adaptability of their octahedral networks. One exceptional structure is fluorite $TiO_2$ that does not exhibit oxygen octahedra network but are known to be stable only under high-pressure conditions. Although identified oxides mostly possess similar oxygen octahedral networks, some fall outside the conventional chemical design space. This highlights the importance of virtual screening conducted without expert knowledge, which can otherwise bias the search space.

## III. CONCLUSIONS AND OUTLOOK

In this study, we propose an approach that enables accurate prediction of the ionic contributions to the dielectric constants ($\boldsymbol{\varepsilon}^{\mathrm{ion}}$), the dominant contributor to the dielectric response in high-κ materials. The key idea is to factorize $\boldsymbol{\varepsilon}^{\text{ion}}$ into BECs and phonon properties, predicting the former using an EGNN and the latter with a highly accurate pretrained MLP. We achieved the accurate predictions of BECs using a training dataset comprising 928 oxides. Compared to conventional single-model approaches, our decomposition and integration strategy leads to a significant improvement in prediction of $\boldsymbol{\varepsilon}^{\text{ion}}$. Using this framework, we successfully identified 31 high-κ oxides from a pool of 8,717 candidates.

We emphasize that, as indicated by Eq. (7) in the *Methods* section, larger $\boldsymbol{\varepsilon}^{\text{ion}}$ are associated with marginally stable optical phonon frequencies. Consequently, the prediction of high dielectric constants becomes highly sensitive to even small errors in phonon frequencies. Moreover, if the phonon frequencies are erroneously predicted to be imaginary, the corresponding dielectric constants cannot be evaluated. Therefore, accurate prediction of phonon properties is a critical aspect of this study. This implies that improvements in both the first-principles calculations and the MLP trained based on their outputs directly contribute to enhancing the accuracy of predicting the $\boldsymbol{\varepsilon}^{\text{ion}}$.

As future directions following this study, two main avenues can be considered: extending the material systems to non-oxides and incorporating finite-temperature effects. The former can be realized simply by expanding the dataset to include non-oxides. On the other hand, the latter can be addressed by performing anharmonic phonons [131] or molecular dynamics calculations [132] using MLP. These extensions will improve the screening of superior high-κ dielectrics. Furthermore, the screened candidates can be theoretically investigated in depth in terms of band alignment with contact metals and defects that affect conductivity and dielectric constants.

Finally, we emphasize that our physics-based factorization approach is applicable not only to dielectric tensors, but also to a broad range of material properties that can be derived from distinct fundamental quantities. We anticipate that this study will offer new perspectives for predicting physical properties.

## Ⅳ. METHODS

### A. Definition and computational approach of force constant matrix

Force constant matrix $\Phi_{j\alpha,j'\beta}$ represents the rate of change of the restoring force that arises in the $\alpha$ direction on atom $j$ when atom $j'$ is slightly displaced in the $\beta$ direction from the equilibrium positions. For atoms $j$ and $j'$, and coordinate components $\alpha$ and $\beta$, the elements of the force constant matrix are defined as:

$$\Phi_{j\alpha,j'\beta} = \frac{\partial^2 E}{\partial u_{j\alpha} \partial u_{j'\beta}} = -\frac{\partial F_{j\alpha}}{\partial u_{j'\beta}}, \tag{4}$$

where $E$ denotes potential energy of the system, and $u_{j\alpha}$ denotes displacement of atom $j$ in the $\alpha$ direction.

In this calculation, the force constant matrix is obtained using finite-difference method, whereby an atom is displaced by 0.01 Å, and the resulting forces on the other atoms are subsequently evaluated using MLPs [133] as follows:

$$\Phi_{j\alpha,j'\beta} \approx -\frac{\Delta F_{j\alpha}}{\Delta u_{j'\beta}}. \tag{5}$$

The training dataset of the MLPs adopted in this study were generated using the PBE functional. Nevertheless, as summarized in Table 1, all the models attain reasonable accuracy of predicting the $\boldsymbol{\varepsilon}^{\text{ion}}$.

## B. Derivation of the ionic dielectric constant in the static limit

The ionic dielectric tensor, which denotes the ionic contribution to the dielectric constants, is expressed by the following expression [30,31]:

$$\varepsilon_{\alpha\beta}^{ion}(\omega) = \frac{4\pi}{V_0}\sum_{m}\frac{\left(\sum_{j\alpha'} Z_{j,\alpha\alpha'}^{*} U_{j\alpha',m\boldsymbol{q}=\boldsymbol{0}}\right)\left(\sum_{j'\beta'} Z_{j',\beta\beta'}^{*} U_{j'\beta',m\boldsymbol{q}=\boldsymbol{0}}\right)}{\omega_m^2 - \omega^2}, \tag{6}$$

where $V_0$ is the volume of the unit cell, and $U_{j\alpha',m\boldsymbol{q}=\boldsymbol{0}} = \frac{e_{m,j\alpha'}}{\sqrt{M_j}}$ denotes the normalized eigen-displacement associated with the dynamical matrix $\boldsymbol{D}$ at the $\Gamma$ point. Taking the static limit ($\omega \to 0$), Eq. (6) simplifies to:

$$\varepsilon_{\alpha\beta}^{ion}(0) = \frac{4\pi}{V_0}\sum_{m}\frac{\left(\sum_{j\alpha'} Z_{j,\alpha\alpha'}^{*} U_{j\alpha',m\boldsymbol{q}=\boldsymbol{0}}\right)\left(\sum_{j'\beta'} Z_{j',\beta\beta'}^{*} U_{j'\beta',m\boldsymbol{q}=\boldsymbol{0}}\right)}{\omega_m^2}. \tag{7}$$

To proceed, we employ the spectral decomposition of the $\boldsymbol{D}$, whose inverse is given by:

$$(\boldsymbol{D}^{-1})_{j\alpha',j'\beta'} = \sum_{m}\frac{e_{m,j\alpha'}e_{m,j'\beta'}}{\omega_m^2}. \tag{8}$$

Using the relation between the normalized displacements and the dynamical matrix:

$$\frac{1}{\sqrt{M_j M_{j'}}}(\boldsymbol{D}^{-1})_{j\alpha',j'\beta'} = \sum_{m}\frac{U_{j\alpha',m\boldsymbol{q}=\boldsymbol{0}}U_{j'\beta',m\boldsymbol{q}=\boldsymbol{0}}}{\omega_m^2}, \tag{9}$$

we can rewrite the ionic dielectric tensor as:

$$\varepsilon_{\alpha\beta}^{ion}(0) = \frac{4\pi}{V_0}\sum_{j\alpha',j'\beta'} Z_{j,\alpha\alpha'}^{*}\left(\frac{1}{\sqrt{M_j M_{j'}}}(\boldsymbol{D}^{-1})_{j\alpha',j'\beta'}\right) Z_{j',\beta\beta'}^{*}. \tag{10}$$

Finally, by introducing the force constant matrix $\boldsymbol{\Phi}$, Eq. (2) is obtained (see Appendix E for details). Note that this expression is valid only when all optical phonon frequencies are real ($\omega_m^2 > 0$). Thus, the presence of imaginary phonon modes, which signal dynamical instability of the lattice, invalidates the above formulation and prevents a meaningful evaluation of the ionic dielectric tensor. In this study, $(\boldsymbol{\Phi}^{-1})_{j\alpha',j'\beta'}$ were evaluated using the Phonopy code [133].

## C. EGNN model

The overall architecture of our ML model is illustrated in Appendix A. The model architecture is inspired by EGNNs such as Nequip, which also leverage symmetry-aware message passing mechanisms [35–37,134]. The model is implemented primarily using the e3nn library [135], which enables equivariant operations under Euclidian symmetry. In this model, node and edge features are represented using irreducible representations of the three-dimensional rotation and inversion group *O*(3). Each representation is labeled by two indices, namely the angular momentum degree (*l* = 0, 1, 2, ...), and the parity, where "e" and "o" stand for even and odd parity, respectively. The nodes are initialized using atomic embedding vectors in the same manner as in CGCNN [136], while the edges are encoded based on interatomic distances via radial basis functions. The initialized node features and edge features are represented as 0e. Additionally, the normalized relative vectors between atoms are encoded into the 0e, 1o, and 2e components using spherical harmonics. Furthermore, by taking tensor products between 0e edge features and corresponding spherical harmonic representations, the edge features are expanded into the 0e, 1o, and 2e components.

Through convolution layers, tensor products are computed between node features and edge features, and the node features ultimately have up to 0e, 1o, 1e, and 2e components. Finally, the model outputs the 0e and 2e components for the dielectric tensor, and the 0e, 1e, and 2e components for the BEC. More details are described in Appendix A.

The EGNN model was trained using specific hyperparameters. The feature dimensions were set to 64 for 0e, 16 for 1e and 1o, and 32 for 2e. The edge dimension was fixed at 32, and three convolutional layers with SiLU activations [137] were used. The batch size was set to 16. Except for the initial learning rate, all hyperparameters were kept fixed through all experiments. The initial learning rate was optimized experimentally, starting from $6.25 \times 10^{-4}$ and doubled incrementally up to a maximum value of $1.60 \times 10^{-1}$. The optimized learning rates determined for each task are summarized in Supplemental Table 1 [34]. For model training, the mean squared error (MSE) was used as the loss function, and optimization was performed using the Adam optimizer [138]. The selected learning rate decayed by a factor of 0.9 every 30 epochs. To ensure reproducibility, a fixed random seed was used throughout all experiments. All training runs were performed on four NVIDIA A100 PCIe GPU with 80 GB of memory.

### D. Training and test datasets

Model evaluation was carried out using 5-fold cross-validation. In each fold, 20 % of the data was reserved for testing, while the remaining 80% was split into 80% for training and 20% for validation. Early stopping was applied based on validation loss, with a maximum of 300 epochs. Training was terminated if no improvement occurred over 100 consecutive epochs.

After cross-validation, the model was retrained on the entire dataset of 928 oxides using the optimized initial learning rate to produce the final predictive model. This step ensured the use of all available data and maximized prediction accuracy. The total number of training epochs in this retraining phase was calculated as the average across the five folds. This final predictive model was then used for virtual screening.

### E. Details of first-principles calculations

First-principles calculations were performed using the projector augmented-wave (PAW) [139] method, as implemented in the Vienna ab initio simulation package (VASP) [140]. Structure relaxation and dielectric tensor calculations were performed using PBEsol+$U$, and band gap calculations were carried out with the DDH. The PAW dataset, the effective $U$ parameters, cutoff energies, densities of the $k$-point mesh, structure optimization criteria were the same as those used in our previous study [38].

## ACKNOWLEDGMENTS

This work has been supported by JST FOREST Program (JPMJFR235S), and KAKENHI (25K01486).

## DATA AVAILABILITY

All data supporting the findings of this study will be made freely accessible at the time of publication.

## CODE AVAILABILITY

The codes used in this study will be made freely accessible at the time of publication.

### Supplemental Material

The Supplemental Material provides comprehensive supporting information for this study. It includes the distributions of the eigenvalues of the BEC tensors, and a systematic comparison of r²SCAN and dielectric-dependent hybrid (DDH) functionals with experimental band-gap data. In addition, the prediction accuracies for dielectric constants are reported. Detailed information on the 31 identified high-$\kappa$ oxides and the nine oxides included in the training dataset is also provided.

# Appendices

## Appendix A: Details of EGNN architecture in this study.

In this study, EGNN architectures were constructed and trained to predict three target properties: BECs, $\boldsymbol{\varepsilon}^{\mathrm{ele}}$, and $\boldsymbol{\varepsilon}^{\mathrm{ion}}$. An overview of the architecture is provided in Fig. 5. The model is implemented using the e3nn [135] library. The SiLU function is adopted as the activation function.

The architecture operates as follows. Nodes and edges are first vectorized, with edges constructed using a cutoff distance of 5 Å. Convolutional operations are then performed over three fixed layers, processing features corresponding 0e, 1o, 1e, and 2e representations. Residual connections, inspired by ResNet [37], are incorporated within the convolution layers to enhance training stability and performance. For the dielectric constants, only the 0e and 2e representations are used in the final output, whereas for the prediction of BECs, the 0e, 1e, and 2e representations are employed, considering the symmetries of the matrices.

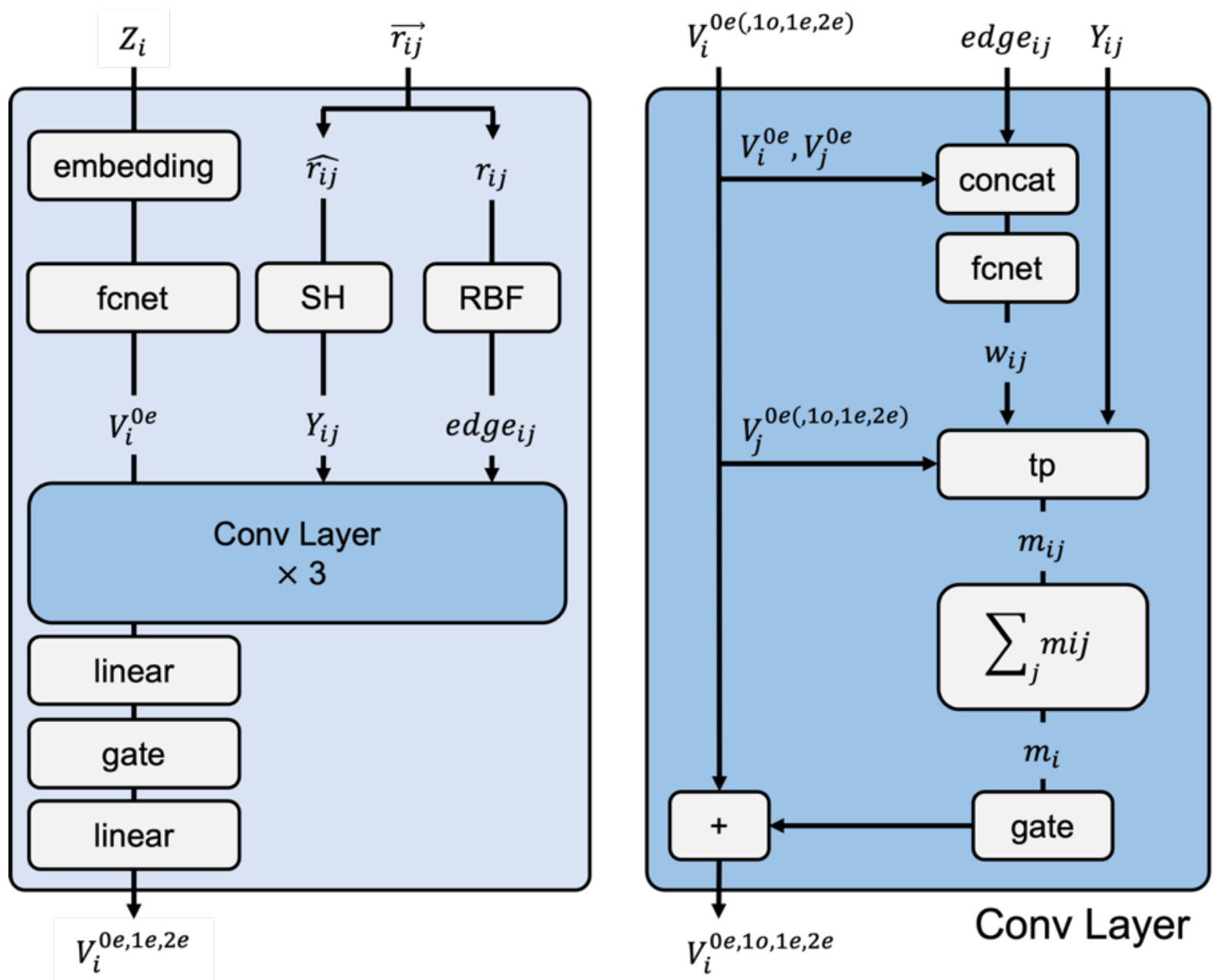

FIG. 5. The EGNN architecture adopted in this study.

The EGNN components are described below.

・ Node Embedding (embedding)：

Node features are constructed from atomic numbers using a dictionary-based approach, employing the same vector representation as used in CGCNN [136]. These vectors are 92-dimensional one-hot encodings incorporating descriptors such as atomic number and electronegativity.

・ Radial Basis Function (RBF)：

The edge vectorization is carried out using the *soft_one_hot_linspace* method from the math module of e3nn, enabling the construction of smooth radial features based on interatomic distance $r_{ij}$ through Gaussian basis functions.

・ Fully Connected Net Layer (fcnet)：

The function $\mathrm{fcnet}(x)$ is implemented using the FullyConnectedNet class from the nn module of e3nn, which is defined as:

$$\mathrm{fcnet}(x) = \frac{xW}{\sqrt{h^{in}\sigma^2}}. \tag{A1}$$

Here, $h^{\mathrm{in}}$ and $h^{\mathrm{out}}$ denote the input and output dimensions, $W \in \mathbb{R}^{h^{\mathrm{in}} \times h^{\mathrm{out}}}$ is a learnable weight matrix, and $\sigma$ represents the standard deviation of the input $x$.

・ Linear Layer（linear）：

This layer combines a fully connected transformation with a nonlinear activation, while maintaining $O(3)$ equivariance. The transformation is formulated as:

$$\mathrm{linear}(x) = \left( \bigoplus_l (W^l \otimes \mathrm{I}_{2l+1}) x^l \right) + b^0. \tag{A2}$$

Here, $\mathrm{I}_{2l+1}$ is the identity matrix in the representation space of degree $l$, $W^l \in \mathbb{R}^{n_l^{out} \times n_l^{in}}$ are learnable weights and $b^0$ is the bias term added only to scalar features. This operation is implemented using the Linear class from the o3 module of e3nn.

・ Equivariant Nonlinear Layer（gate）：

This layer applies non-linear transformations to both scalar ($l = 0$) and higher-order ($l > 0$) features. The operation is defined as:

$$\mathrm{gate}(x, g, y) = \left( \bigoplus_i \phi(x_i) \right) \oplus \left( \bigoplus_j \phi(g_j) y_j \right). \tag{A3}$$

Here, $\phi$ denotes the activation function, $x_i$ and $g_j$ are scalar features, and $y_j$ represents higher-order features. The scalar inputs $x_i$ and $g_j$ are typically obtained by splitting the input scalar features into two parts. This component is implemented via the Gate class from the nn module of e3nn.

・ Tensor Product（tp）：

In our model, tensor product operations are employed to lift scalar features into higher-order representations such as vectors and second-rank tensors, while preserving spatial equivariance. This enables the network to encode structural information in a symmetry-aware manner.

Specifically, the tensor product is computed between node-assigned features and spherical harmonics, which serve as irreducible representations of the $O(3)$ group and capture angular dependencies. For instance, taking the tensor product of a scalar feature and a spherical harmonic allows the model to construct features of higher angular momentum (i.e., $l > 0$).

The resulting features are decomposed into components corresponding to their total angular momentum $l$. When combining two features of angular momenta $l_1$ and $l_2$, the output contains components with $|l_1 - l_2| \leq l \leq l_1 + l_2$.

In this work, the model is designed to handle components up to $l = 2$. Through this mechanism, the model effectively incorporates geometric information such as interatomic distances and directions while maintaining compliance with spatial symmetries. As a result, it naturally constructs representations that are invariant or equivariant under transformations like rotation and reflection.

The tensor product operation is formulated in e3nn as:

$$z_w = \mathrm{tp}(x, y, w) = \sum_{u,v} w_{uvw} x_u \otimes y_v \tag{A4}$$

Here, $x_u$ represents the features assigned to node $j$ (denoted $V_j$), $y_v$ corresponds to the spherical harmonic $Y_{ij}$ and $w_{uvw}$ are weights derived from the edge features $w_{ij}$ . This operation is implemented efficiently via the FullyConnectedTensorProduct class within o3 module of the e3nn library.

・ Convolution Layer（Conv Layer）：

Our model primarily updates atomic features using operations based on tensor products. The tensor product weights $w_{ij}$ are generated from the vectorized edge features (denoted $edge_{ij}$) and scalar features of atom $i$ and $j$, denoted $V_I{}^{l=0}$ and $V_j{}^{l=0}$:

$$w_{ij} = \mathrm{fcnet}\left(\mathrm{concatenate}\left(V_i{}^{l=0}, V_j{}^{l=0}, edge_{ij}\right)\right) \tag{A5}$$

Using the generated weight $w_{ij}$, the feature vector $V_I$ associated with atom $i$ is updated. Inspired by the ResNet architecture [37], a residual connection is employed to facilitate efficient gradient propagation:

$$V_i^{(t+1)} = V_i^{(t)} + \sum_j \mathrm{tp}\left(V_j, Y_{ij}, w_{ij}\right) \tag{A6}$$

Here, $t$ is the layer index, and $Y_{ij}$ denotes the spherical harmonics computed based on the relative positions of atom $i$ and $j$.

## Appendix B: Previous studies on machine-learning prediction of Born effective charges.

Here, we summarize previous studies on the ML prediction of the BECs.

- Falletta *et al.* [39] (2025) constructed MLP-based models to simultaneously predict energies, forces, polarization, polarizability, and BECs from differential relationships between a generalized potential function and applied external fields. Models were trained for $\alpha$-$SiO_2$ and $BaTiO_3$ using *ab initio* molecular dynamics (AIMD) snapshots independently. The resulting MAEs for BECs are 0.005 for $\alpha$-$SiO_2$ and 0.011 for $BaTiO_3$.

- Shimizu *et al.* [41] (2023) developed a neural-network model based on vector atomic fingerprints for predicting BECs from atomic structures to study ion migration under electric fields. The model was trained on BECs from 17,900 $Li_3PO_4$ structures, including both pristine and defect-containing configurations. The model achieved an RMSE of 0.038.
- Kutana *et al*. [42] (2025) predicted the BECs in perovskite oxides ($ABO_3$ in the *Pnma* space group; $A$ =Ca, Sr, Ba, Pb and $B$=Ti, Zr, Hf) as well as those in $Li_3PO_4$ and $ZrO_2$ using EGNNs. The reported MAEs for BECs are 0.020–0.032 when trained jointly and 0.009-0.033 when trained separately for each material class.
- Schmiedmayer and Kresse [40] (2024) developed a tensorial equivariant derivative learning that predict BECs using $\lambda$-SOAP descriptors combined with linear regression for predicting finite-temperature infrared spectra. The resulting RMSEs were 0.032 for liquid water and 0.040 for $MAPbI_3$.

Although the numerical errors reported in previous studies [38–41] are approximately one order of magnitude smaller than those obtained with the present model (MAE = 0.231 ± 0.018, RMSE = 0.373 ± 0.042), these approaches are generally applied to specific materials or limited classes of materials.

## Appendix C: Acoustic sum rule for predicted Born effective charges.

The BEC tensors satisfy the acoustic sum rule:

$$\sum_j Z^*_{j,\alpha\beta} = 0, \tag{C1}$$

because rigid translation of the crystal does not modify the internal charge distribution, so the macroscopic polarization must remain unchanged.

$$\Delta\mathcal{P}_\alpha = \sum_j \frac{\partial P_\alpha}{\partial u_{j\beta}} u_{j\beta} = \frac{1}{V_0}\left(\sum_j Z^*_{j,\alpha\beta}\right) u_\beta = 0. \tag{C2}$$

In the present work, this constraint is not explicitly imposed during the training or inference of the EGNN for predicting the BECs. To assess its impact, we quantitatively evaluated the degree to which the acoustic sum rule is satisfied. Specifically, we introduce direction-resolved neutrality metrics that characterize the degree of electrical neutrality along each Cartesian direction $\alpha = x, y, z$, defined as

$$\Delta = \frac{1}{3}, \tag{C3}$$

where

$$\overline{Z^*_{\alpha\alpha}} = \frac{1}{N}\left|\sum_{j=1}^{N} Z^*_{j,\alpha\alpha}\right|. \tag{C4}$$

These metrics measure the residual net effective charge along each direction, and thus provide a measure of the accuracy with which electrical neutrality is satisfied. Across all oxides in all test sets of the 5-fold cross-validation, we obtain $\Delta = 0.061 \pm 0.060$, indicating that the relative residual violation is typically small compared with the absolute value of the nominal oxygen charge of −2.

In addition, we explicitly enforced the acoustic sum rule on the predicted $\boldsymbol{\varepsilon}^{\mathrm{ion}}$ by constantly shifting the BECs of all atoms so that the ASR is satisfied as:

$$Z_{j,\alpha\beta}^{\mathrm{corr},*} = Z_{j,\alpha\beta}^{*} - \frac{1}{N}\sum_{j'} Z_{j',\alpha\beta}^{*}, \tag{C5}$$

where $N$ donates the number of atoms in the cell.

With the original $Z_j^*$, RMSE is $0.099 \pm 0.012$ and $R^2$ is $0.911 \pm 0.017$, while with the $Z_j^{\mathrm{corr},*}$, RMSE is $0.097 \pm 0.012$ and $R^2$ is $0.914 \pm 0.017$. Here, the mean values and standard deviations are evaluated over the five test sets of the 5-fold cross-validation used for BEC training, indicating that explicit enforcing of the acoustic sum rule has no measurable impact on the prediction accuracy of the $\boldsymbol{\varepsilon}^{\mathrm{ion}}$.

## Appendix D: Previous studies on machine-learning prediction of dielectric constants.

Several studies have reported ML-based predictions of the electronic, ionic, and total dielectric constants. The dielectric constant is a second-rank tensor, and its spherically average is obtained by averaging the diagonal components. In evaluating the prediction accuracies of tensors, the eigenvalues are utilized for assessment. Here, previous studies are summarized, and the accuracy metrics are tabulated in Table 4.

- Takahashi *et al*. (2020) [23] employed a random forest (RF) model with compositional and structural descriptors. They predicted spherically averaged electronic and ionic dielectric constants ($\varepsilon_{\mathrm{ave}}^{\mathrm{ele}}$ and $\varepsilon_{\mathrm{ave}}^{\mathrm{ion}}$) across 1,266 oxides.
- Morita *et al*. (2020) [27] reported ML models for predicting only $\varepsilon_{\mathrm{ave}}^{\mathrm{ele}}$ in 1,364 crystalline materials. They benchmarked a descriptor-based support vector regression (SVR) model against a crystal-graph MEGNet [141] model with transfer learning.
- Kim *et al.* (2022) [26] constructed a gradient boosting regression (GBR) model for predicting dielectric properties using first-principles data of 7,113 inorganic crystals curated in the MPD [53].
- Shimano *et al*. (2023) [25] employed RF and graph convolutional neural network (SchNet [29]) models to predict the dielectric constants of 2,150 oxides derived from elemental substitutions in rutile, perovskite, Ruddlesden–Popper, and double-perovskite prototypes.
- Hu *et al*. (2024) [24] predicted the averaged electronic and ionic dielectric constants of 722 binary and ternary oxides using SVR models incorporating detailed structural features. Although their prediction accuracy improves upon that of Takahashi *et al*. [23], this is likely due to the restriction of the compositional space to binary and ternary oxides.
- Mao *et al*. (2024) [18] applied an EGNN combined with transfer learning from an MLP, targeting 6,648 inorganic crystals, whose dielectric tensor data were collected from the MPD. They limited the target crystals to those for which all components of the dielectric tensors are below 100. Only the prediction accuracies for the summed dielectric constants were reported.
- Lou *et al*. (2024) [28] proposed an EGNN (AnisoNet) to directly predict the $\varepsilon^{\mathrm{ele}}$ from crystal structures. The model was trained and evaluated on 6,706 inorganic crystals registered in the MPD, with the dataset restricted to entries for which all diagonal components ($\varepsilon_{\mathrm{ii}}$) satisfy $\varepsilon_{\mathrm{ii}} \geq 1$. In addition, entries with $\varepsilon_{\mathrm{ave}}^{\mathrm{ele}} \geq 15$ were excluded.

They achieved MAE = 0.311 for $\varepsilon_{\mathrm{ave}}^{\mathrm{ele}}$ in a base-10 logarithmic scale. As the $R^2$ was not reported, it is not included in Table 4.

TABLE 4. $R^2$ of the prediction of dielectric constants in previous studies and this study. There are cases where only spherically averaged quantities are predicted, as well as cases where eigenvalues are predicted. “--” indicate the absence of the data. “Scale” refers to whether the evaluation metrics are computed on a linear or base-10 logarithmic scale.

| **Study** | **Dataset** | **Model** | Scale | $\varepsilon_{\mathrm{ave}}^{\mathrm{ele}}$ | $\varepsilon_{\mathrm{ave}}^{\mathrm{ion}}$ | $\varepsilon_{\mathrm{ave}}$ | $\lambda_{\boldsymbol{\varepsilon}^{\mathrm{ele}}}$ | $\lambda_{\boldsymbol{\varepsilon}^{\mathrm{ion}}}$ | $\lambda_{\boldsymbol{\varepsilon}}$ |
|---|---|---|---|---|---|---|---|---|---|
| Takahashi *et al.* (2020) | 1,266 oxides | RF | $\log_{10}$ | 0.89 | 0.73 | -- | -- | -- | -- |
| Morita *et al.* (2020) | 1,364 inorganic crystals | SVR | Linear | 0.86 | -- | -- | -- | -- | -- |
| | | MEGNet | Linear | 0.84 | -- | -- | -- | -- | -- |
| Kim *et al.* (2022) | 7,113 inorganic crystals | GBR | $\log_{10}$ | 0.83 | 0.67 | -- | -- | -- | -- |
| Shimano *et al.* (2023) | 2,150 oxides in limited structural frameworks | RF | $\log_{10}$ | 0.92 | 0.89 | -- | -- | -- | -- |
| | | SchNet | $\log_{10}$ | -- | 0.85 | -- | -- | -- | -- |
| Hu *et al.* (2024) | 722 binary and ternary oxides | SVR | $\log_{10}$ | 0.94 | 0.79 | 0.89 | -- | -- | -- |
| Mao *et al.* (2024) | 6648 inorganic crystals | EGNN | $\log_{10}$ | -- | -- | 0.89 | -- | -- | 0.78 |
| This study | 928 oxides for $\boldsymbol{\varepsilon}^{\mathrm{ele}}$<br>839 oxides for $\boldsymbol{\varepsilon}^{\mathrm{ion}}$ and $\boldsymbol{\varepsilon}$ | EGNN, Joint model | $\log_{10}$ | 0.74 | 0.91 | 0.91 | 0.71 | 0.91 | 0.91 |

## Appendix E: Formulation and properties of dynamical matrix.

The potential energy $E$ of a system can be expressed using the force constant matrix $\Phi_{j\alpha,j'\beta}$ as follows:

$$E = \frac{1}{2}\sum_{j\alpha,j'\beta} u_{j\alpha}\Phi_{j\alpha,j'\beta}u_{j'\beta}\,, \tag{E1}$$

where $u_{j\alpha}$ is the displacement of atom $j$ in the $\alpha$ direction. Introducing the mass-normalized displacement $\tilde{u}_{j\alpha} = \sqrt{M_j}u_{j\alpha}$, where $M_j$ is the mass of atom $j$, Eq. (F1) can be reformulated as:

$$E = \frac{1}{2}\sum_{j\alpha,j'\beta} \tilde{u}_{j\alpha}\left(\frac{1}{\sqrt{M_j}}\Phi_{j\alpha,j'\beta}\frac{1}{\sqrt{M_{j'}}}\right)\tilde{u}_{j'\beta}\,. \tag{E2}$$

Defining the reduced force constant matrix as $\tilde{\Phi}_{j\alpha,j'\beta} = \frac{1}{\sqrt{M_j}}\Phi_{j\alpha,j'\beta}\frac{1}{\sqrt{M_{j'}}}$, the symmetric matrix $\tilde{\boldsymbol{\Phi}}$ can be diagonalized according to

$$\tilde{\boldsymbol{\Phi}} = \boldsymbol{Q}\boldsymbol{\Omega}^2\boldsymbol{Q}^{\mathrm{T}}, \tag{E3}$$

where $\boldsymbol{\Omega}^2 = \mathrm{diag}(\cdots, \omega_m^2, \cdots)$, $\boldsymbol{Q}$ is an orthogonal matrix whose columns are the eigenvectors $\boldsymbol{e}_m$ corresponding to the phonon modes indexed by $m$.

The associated eigenvalue problem of Eq. (F3) is given explicitly by:

$$\sum_{j'\beta} \widetilde{\Phi}_{j\alpha,j'\beta}\, e_{m,j'\beta} = \omega_m^2 e_{m,j\alpha}. \tag{E4}$$

By the Blöch theorem, the eigenvector $e_{m,i\alpha}$ of a phonon mode with wave vector $\boldsymbol{q}$ take the form [142,143]:

$$e_{m,j\alpha}(\boldsymbol{q}) = W_{m,j\alpha} \exp\left(\boldsymbol{iq} \cdot \boldsymbol{R}_j\right), \tag{E5}$$

where $W_{m,i\alpha}$ is a periodic function of the crystal, and $\boldsymbol{R}_j$ denotes the position of atom $j$. Substituting Eq. (F5) into Eq. (F4), multiplying both sides by $\exp\left(-\boldsymbol{iq} \cdot \boldsymbol{R}_j\right)$, yields:

$$\sum_{j'\beta} D_{j\alpha,j'\beta}(\boldsymbol{q})\, W_{m,j'\beta} = \omega_m^2 W_{m,j\alpha}, \tag{E6}$$

where the dynamical matrix $D_{j\alpha,j'\beta}(\boldsymbol{q})$ is defined as:

$$\begin{aligned} D_{j\alpha,j'\beta}(\boldsymbol{q}) &= \exp\left(-\boldsymbol{iq} \cdot \boldsymbol{R}_j\right) \widetilde{\Phi}_{j\alpha,j'\beta} \exp\left(\boldsymbol{iq} \cdot \boldsymbol{R}_{j'}\right) \\ &= \frac{1}{\sqrt{M_j M_{j'}}} \Phi_{j\alpha,j'\beta} \exp\left(\boldsymbol{iq} \cdot \left(\boldsymbol{R}_{j'} - \boldsymbol{R}_j\right)\right). \end{aligned} \tag{E7}$$

At the Γ point, i.e., when $\boldsymbol{q} = \boldsymbol{0}$, Eq. (F5) simplifies to:

$$e_{m,j\alpha} = W_{m,j\alpha}, \tag{E8}$$

And the dynamical matrix elements become:

$$D_{j\alpha,j'\beta}(\boldsymbol{0}) = \frac{1}{\sqrt{M_j M_{j'}}} \Phi_{j\alpha,j'\beta} = \widetilde{\Phi}_{j\alpha,j'\beta}. \tag{E9}$$

Based on this formulation, it is possible to construct the force constant matrix from the potential energy $E$ and atomic forces predicted by MLPs. Subsequently, the dynamical matrix can be computed, and through its diagonalization, phonon frequencies and eigen-displacement vectors are obtained.

[1] R. P. Ortiz, A. Facchetti, and T. J. Marks, High-κ Organic, Inorganic, and Hybrid Dielectrics for Low-Voltage Organic Field-Effect Transistors, Chem. Rev. **110**, 205 (2010).

[2] B. Wang, W. Huang, L. Chi, M. Al-Hashimi, T. J. Marks, and A. Facchetti, High-κ Gate Dielectrics for Emerging Flexible and Stretchable Electronics, Chem. Rev. **118**, 5690 (2018).

[3] A. I. Kingon, J.-P. Maria, and S. K. Streiffer, Alternative dielectrics to silicon dioxide for memory and logic devices, Nature **406**, 1032 (2000).

[4] G. D. Wilk, R. M. Wallace, and J. M. Anthony, High-κ gate dielectrics: Current status and materials properties considerations, J. Appl. Phys. **89**, 5243 (2001).

[5] J. Robertson, High dielectric constant oxides, The European Physical Journal Applied Physics **28**, 265 (2004).

[6] C. H. Ahn, K. M. Rabe, and J.-M. Triscone, Ferroelectricity at the Nanoscale: Local Polarization in Oxide Thin Films and Heterostructures, Science **303**, 488 (2004).

[7] M. Esro, O. Kolosov, P. J. Jones, W. I. Milne, and G. Adamopoulos, Structural and Electrical Characterization of $SiO_2$ Gate Dielectrics Deposited from Solutions at Moderate Temperatures in Air, ACS Appl. Mater. Interfaces **9**, 529 (2017).

[8] K. Eisenbeiser et al., Field effect transistors with $SrTiO_3$ gate dielectric on Si, Appl. Phys. Lett. **76**, 1324 (2000).

[9] O. Trithaveesak, J. Schubert, and Ch. Buchal, Ferroelectric properties of epitaxial $BaTiO_3$ thin films and heterostructures on different substrates, J. Appl. Phys. **98**, (2005).

[10] D. G. Schlom, L. Chen, X. Pan, A. Schmehl, and M. A. Zurbuchen, A Thin Film Approach to Engineering Functionality into Oxides, Journal of the American Ceramic Society **91**, 2429 (2008).

[11] S. M. George, Atomic Layer Deposition: An Overview, Chem. Rev. **110**, 111 (2010).

[12] J. Robertson, High dielectric constant gate oxides for metal oxide Si transistors, Reports on Progress in Physics **69**, 327 (2006).

[13] M. Balog, M. Schieber, M. Michman, and S. Patai, Chemical vapor deposition and characterization of $HfO_2$ films from organo-hafnium compounds, Thin Solid Films **41**, 247 (1977).

[14] K. Yim, Y. Yong, J. Lee, K. Lee, H.-H. Nahm, J. Yoo, C. Lee, C. Seong Hwang, and S. Han, Novel high-κ dielectrics for next-generation electronic devices screened by automated ab initio calculations, NPG Asia Mater. **7**, e190 (2015).

[15] I. Petousis, D. Mrdjenovich, E. Ballouz, M. Liu, D. Winston, W. Chen, T. Graf, T. D. Schladt, K. A. Persson, and F. B. Prinz, High-throughput screening of inorganic compounds for the discovery of novel dielectric and optical materials, Sci. Data **4**, 160134 (2017).

[16] A. Gopakumar, K. Pal, and C. Wolverton, Identification of high-dielectric constant compounds from statistical design, NPJ Comput. Mater. **8**, 146 (2022).

[17] J. Riebesell, T. W. Surta, R. E. A. Goodall, M. W. Gaultois, and A. A. Lee, Discovery of high-performance dielectric materials with machine-learning-guided search, Cell Rep. Phys. Sci. **5**, 102241 (2024).

[18] Z. Mao, W. Li, and J. Tan, Dielectric tensor prediction for inorganic materials using latent information from preferred potential, NPJ Comput. Mater. **10**, 265 (2024).

[19] R. Ruff, P. Reiser, J. Stühmer, and P. Friederich, Connectivity optimized nested line graph networks for crystal structures, Digital Discovery **3**, 594 (2024).

[20] S. S. Omee, S.-Y. Louis, N. Fu, L. Wei, S. Dey, R. Dong, Q. Li, and J. Hu, Scalable deeper graph neural networks for high-performance materials property prediction, Patterns **3**, 100491 (2022).

[21] K. Choudhary and B. DeCost, Atomistic Line Graph Neural Network for improved materials property predictions, NPJ Comput. Mater. **7**, 185 (2021).

[22] M. Gajdoš, K. Hummer, G. Kresse, J. Furthmüller, and F. Bechstedt, Linear optical properties in the projector-augmented wave methodology, Phys. Rev. B **73**, 045112 (2006).

[23] A. Takahashi, Y. Kumagai, J. Miyamoto, Y. Mochizuki, and F. Oba, Machine learning models for predicting the dielectric constants of oxides based on high-throughput first-principles calculations, Phys. Rev. Mater. **4**, 103801 (2020).

[24] Y. Hu et al., Accurate prediction of dielectric properties and bandgaps in materials with a machine learning approach, Appl. Phys. Lett. **125**, (2024).

[25] Y. Shimano, A. Kutana, and R. Asahi, Machine learning and atomistic origin of high dielectric permittivity in oxides, Sci. Rep. **13**, (2023).

[26] E. Kim, J. Kim, and K. Min, Prediction of dielectric constants of $ABO_3$-type perovskites using machine learning and first-principles calculations, Physical Chemistry Chemical Physics **24**, 7050 (2022).

[27] K. Morita, D. W. Davies, K. T. Butler, and A. Walsh, Modeling the dielectric constants of crystals using machine learning, Journal of Chemical Physics **153**, (2020).

[28] Y. Lou and A. M. Ganose, Discovery of highly anisotropic dielectric crystals with equivariant graph neural networks, Faraday Discuss. **256**, 255 (2024).

[29] K. T. Schütt, H. E. Sauceda, P. J. Kindermans, A. Tkatchenko, and K. R. Müller, SchNet - A deep learning architecture for molecules and materials, Journal of Chemical Physics **148**, (2018).

[30] X. Gonze and C. Lee, Dynamical matrices, Born effective charges, dielectric permittivity tensors, and interatomic force constants from density-functional perturbation theory, Phys. Rev. B **55**, 10355 (1997).

[31] A. A. Maradudin, E. W. Montroll, and G. H. Weiss, *Theory of Lattice Dynamics in the Harmonic Approximation*, 2nd ed., Vol. 3 (Academic Press, New York, 1971).

[32] Ph. Ghosez, J.-P. Michenaud, and X. Gonze, Dynamical atomic charges: The case of $ABO_3$ compounds, Phys. Rev. B **58**, 6224 (1998).

[33] R. Resta, Macroscopic polarization in crystalline dielectrics: the geometric phase approach, Rev. Mod. Phys. **66**, 899 (1994).

[34] See Supplemental Material [URL] for providing the distributions of the eigenvalues of the BEC tensors, and a systematic comparison of $r^2$SCAN and dielectric-dependent hybrid (DDH) functionals with experimental band-gap data, the prediction accuracies for dielectric constants, detailed information on the 31 identified high-κ oxides and the nine oxides included in the training dataset.

[35] S. Batzner, A. Musaelian, L. Sun, M. Geiger, J. P. Mailoa, M. Kornbluth, N. Molinari, T. E. Smidt, and B. Kozinsky, E(3)-equivariant graph neural networks for data-efficient and accurate interatomic potentials, Nat. Commun. **13**, 2453 (2022).

[36] N. Thomas, T. Smidt, S. Kearnes, L. Yang, L. Li, K. Kohlhoff, and P. Riley, Tensor field networks: Rotation- and translation-equivariant neural networks for 3D point clouds, ArXiv Preprint ArXiv:1802.08219 (2018).

[37] K. He, X. Zhang, S. Ren, and J. Sun, *Deep Residual Learning for Image Recognition*, in *2016 IEEE Conference on Computer Vision and Pattern Recognition (CVPR)* (IEEE, 2016), pp. 770–778.

[38] Y. Kumagai, N. Tsunoda, A. Takahashi, and F. Oba, Insights into oxygen vacancies from high-throughput first-principles calculations, Phys. Rev. Mater. **5**, 123803 (2021).

[39] S. Falletta, A. Cepellotti, A. Johansson, C. W. Tan, M. L. Descoteaux, A. Musaelian, C. J. Owen, and B. Kozinsky, Unified differentiable learning of electric response, Nat. Commun. **16**, 4031 (2025).

[40] B. Schmiedmayer and G. Kresse, Derivative learning of tensorial quantities—Predicting finite temperature infrared spectra from first principles, Journal of Chemical Physics **161**, (2024).

[41] K. Shimizu, R. Otsuka, M. Hara, E. Minamitani, and S. Watanabe, Prediction of Born effective charges using neural network to study ion migration under electric fields: applications to crystalline and amorphous $Li_3PO_4$, Science and Technology of Advanced Materials: Methods **3**, (2023).

[42] A. Kutana, K. Shimizu, S. Watanabe, and R. Asahi, Representing Born effective charges with equivariant graph convolutional neural networks, Sci. Rep. **15**, 16719 (2025).

[43] Max. Born and Kun. Huang, *Dynamical Theory of Crystal Lattices* (Clarendon Press, Oxford, 1998).

[44] X. Fu, B. M. Wood, L. Barroso-Luque, D. S. Levine, M. Gao, M. Dzamba, and C. L. Zitnick, Learning Smooth and Expressive Interatomic Potentials for Physical Property Prediction, ArXiv Preprint ArXiv:2502.12147 (2025).

[45] B. Rhodes, S. Vandenhaute, V. Šimkus, J. Gin, J. Godwin, T. Duignan, and M. Neumann, Orb-v3: atomistic simulation at scale, ArXiv Preprint ArXiv:2504.06231 (2025).

[46] I. Batatia et al., A foundation model for atomistic materials chemistry, ArXiv Preprint ArXiv:2401.00096 (2024).

[47] J. Kim, J. Kim, J. Kim, J. Lee, Y. Park, Y. Kang, and S. Han, Data-Efficient Multifidelity Training for High-Fidelity Machine Learning Interatomic Potentials, J. Am. Chem. Soc. **147**, 1042 (2025).

[48] A. Loew, D. Sun, H.-C. Wang, S. Botti, and M. A. L. Marques, Universal machine learning interatomic potentials are ready for phonons, NPJ Comput. Mater. **11**, 178 (2025).

[49] Y. Park, J. Kim, S. Hwang, and S. Han, Scalable Parallel Algorithm for Graph Neural Network Interatomic Potentials in Molecular Dynamics Simulations, J. Chem. Theory Comput. **20**, 4857 (2024).

[50] J. Riebesell, R. E. A. Goodall, P. Benner, Y. Chiang, B. Deng, G. Ceder, M. Asta, A. A. Lee, A. Jain, and K. A. Persson, A framework to evaluate machine learning crystal stability predictions, Nat. Mach. Intell. **7**, 836 (2025).

[51] J. P. Perdew, K. Burke, and M. Ernzerhof, Generalized Gradient Approximation Made Simple, Phys. Rev. Lett. **77**, 3865 (1996).

[52] J. P. Perdew, A. Ruzsinszky, G. I. Csonka, O. A. Vydrov, G. E. Scuseria, L. A. Constantin, X. Zhou, and K. Burke, Restoring the Density-Gradient Expansion for Exchange in Solids and Surfaces, Phys. Rev. Lett. **100**, 136406 (2008).

[53] A. Jain et al., Commentary: The Materials Project: A materials genome approach to accelerating materials innovation, APL Mater. **1**, (2013).

[54] J. H. Skone, M. Govoni, and G. Galli, Self-consistent hybrid functional for condensed systems, Phys. Rev. B **89**, 195112 (2014).

[55] P. B. Jamieson and S. C. Abrahams, Re-examination of the crystal structure of ferroelectric tetragonal bronze-type $Ba_6Ti_2Nb_8O_{30}$, Acta Crystallogr. B **24**, 984 (1968).

[56] R. Someya, Y. J. Shan, and K. Tezuka, Crystal structure and dielectric properties of $Ba_{6-x}Ca_xTi_2Nb_8O_{30}$ tungsten bronze ceramics, Journal of the Ceramic Society of Japan **131**, 22170 (2023).

[57] Y. Yuan, X. M. Chen, and Y. J. Wu, Diffused ferroelectrics of $Ba_6Ti_2Nb_8O_{30}$ and $Sr_6Ti_2Nb_8O_{30}$ with filled tungsten-bronze structure, J. Appl. Phys. **98**, (2005).

[58] G. L. Roberts, R. J. Cava, W. F. Peck, and J. J. Krajewski, Dielectric properties of barium titanium niobates, J. Mater. Res. **12**, 526 (1997).

[59] Y. Itoh, S. Miyazawa, T. Yamada, and H. Iwasaki, Dielectric Property of $Ba_6Ti_2Nb_8O_{30}$ Single Crystal, Jpn. J. Appl. Phys. **9**, 157 (1970).

[60] K. Praveena and K. B. R. Varma, Ferroelectric and optical properties of $Ba_5Li_2Ti_2Nb_8O_{30}$ ceramics potential for memory applications, Journal of Materials Science: Materials in Electronics **25**, 3103 (2014).

[61] K. E. Johnston, C. C. Tang, J. E. Parker, K. S. Knight, P. Lightfoot, and S. E. Ashbrook, The Polar Phase of $NaNbO_3$: A Combined Study by Powder Diffraction, Solid-State NMR, and First-Principles Calculations, J. Am. Chem. Soc. **132**, 8732 (2010).

[62] H. Xu, Y. Su, M. Lou Balmer, and A. Navrotsky, A New Series of Oxygen-Deficient Perovskites in the $NaTi_xNb_{1-x}O_{3-0.5x}$ System: Synthesis, Crystal Chemistry, and Energetics, Chemistry of Materials **15**, 1872 (2003).

[63] A. C. Sakowski-Cowley, K. Łukaszewicz, and H. D. Megaw, The structure of sodium niobate at room temperature, and the problem of reliability in pseudosymmetric structures, Acta Crystallogr. B **25**, 851 (1969).

[64] S. K. Mishra, N. Choudhury, S. L. Chaplot, P. S. R. Krishna, and R. Mittal, Competing antiferroelectric and ferroelectric interactions in $NaNbO_3$: Neutron diffraction and theoretical studies, Phys. Rev. B **76**, 024110 (2007).

[65] D. O. Mishchuk, O. I. V'yunov, O. V. Ovchar, and A. G. Belous, Structural and dielectric properties of solid solutions of sodium niobate in lanthanum and neodymium niobates, Inorganic Materials **40**, 1324 (2004).

[66] Y. J. Shan, H. Mori, R. Wang, W. Luan, H. Imoto, Mitsuruitoh, and T. Nakamura, Powder X-ray diffraction study of ferroelectric phase transition in perovskite oxide $CdTiO_3$, Ferroelectrics **259**, 85 (2001).

[67] R. Caracas and R. E. Cohen, Prediction of a new phase transition in $Al_2O_3$ at high pressures, Geophys. Res. Lett. **32**, 1 (2005).

[68] B. J. Kennedy, Q. Zhou, and M. Avdeev, The ferroelectric phase of $CdTiO_3$: A powder neutron diffraction study, J. Solid State Chem. **184**, 2987 (2011).

[69] S. Sasaki, C. T. Prewitt, J. D. Bass, and W. A. Schulze, Orthorhombic perovskite $CaTiO_3$ and $CdTiO_3$: structure and space group, Acta Crystallogr. C **43**, 1668 (1987).

[70] H. F. Kay and J. L. Miles, The structure of cadmium titanate and sodium tantalate, Acta Crystallogr. **10**, 213 (1957).

[71] T. Sugai and M. Wada, Single Crystal Growth and Some Properties of $Cd_2Ti_2O_5F_2$ and $CdTiO_3$, Jpn. J. Appl. Phys. **18**, 1709 (1979).

[72] A. Garbout, N. Kallel- Kchaou, and M. Férid, Relationship between the structural characteristics and photoluminescent properties of $LnEuTi_2O_7$ (Ln=Gd and Y) pyrochlores, J. Lumin. **169**, 359 (2016).

[73] L. Kong, Z. Zhang, M. de los Reyes, I. Karatchevtseva, G. R. Lumpkin, G. Triani, and R. D. Aughterson, Soft chemical synthesis and structural characterization of $Y_2Hf_xTi_{2-x}O_7$, Ceram. Int. **41**, 5309 (2015).

[74] F. Matteucci, G. Cruciani, M. Dondi, G. Baldi, and A. Barzanti, Crystal structural and optical properties of Cr-doped $Y_2Ti_2O_7$ and $Y_2Sn_2O_7$ pyrochlores, Acta Mater. **55**, 2229 (2007).

[75] O. Knop, F. Brisse, and L. Castelliz, Pyrochlores. V. Thermoanalytic, X-ray, neutron, infrared, and dielectric studies of $A_2Ti_2O_7$ titanates, Can. J. Chem. **47**, 971 (1969).

[76] B. J. Sun, Q. L. Liu, J. K. Liang, J. B. Li, L. N. Ji, J. Y. Zhang, Y. H. Liu, and G. H. Rao, Subsolidus phase relations in the ternary system $SnO_2$–$TiO_2$–$Y_2O_3$, J. Alloys Compd. **455**, 265 (2008).

[77] M. Glerup, O. F. Nielsen, and F. W. Poulsen, The Structural Transformation from the Pyrochlore Structure, $A_2B_2O_7$, to the Fluorite Structure, $AO_2$, Studied by Raman Spectroscopy and Defect Chemistry Modeling, J. Solid State Chem. **160**, 25 (2001).

[78] J. M. Farmer, L. A. Boatner, B. C. Chakoumakos, M.-H. Du, M. J. Lance, C. J. Rawn, and J. C. Bryan, Structural and crystal chemical properties of rare-earth titanate pyrochlores, J. Alloys Compd. **605**, 63 (2014).

[79] T. Ting-ting, W. Li-xi, and Z. Qi-tu, Study on the composite and properties of $Y_2O_3$–$TiO_2$ microwave dielectric ceramics, J. Alloys Compd. **486**, 606 (2009).

[80] B. Muktha, M. H. Priya, G. Madras, and T. N. Guru Row, Synthesis, Structure, and Photocatalysis in a New Structural Variant of the Aurivillius Phase: $LiBi_4M_3O_{14}$ (M = Nb, Ta), J. Phys. Chem. B **109**, 11442 (2005).

[81] M. Bharathy, A. H. Fox, S. J. Mugavero, and H.-C. zur Loye, Crystal growth of inter-lanthanide $LaLn'O_3$ (Ln′=Y, Ho–Lu) perovskites from hydroxide fluxes, Solid State Sci. **11**, 651 (2009).

[82] E. A. Zhurova, Y. Ivanov, V. Zavodnik, and V. Tsirelson, Electron density and atomic displacements in $KTaO_3$, Acta Crystallogr. B **56**, 594 (2000).

[83] A. Tkach, P. M. Vilarinho, and A. Almeida, Role of initial potassium excess on the properties of potassium tantalate ceramics, J. Eur. Ceram. Soc. **31**, 2303 (2011).

[84] O. Aktas, S. Crossley, M. A. Carpenter, and E. K. H. Salje, Polar correlations and defect-induced ferroelectricity in cryogenic $KTaO_3$, Phys. Rev. B **90**, 165309 (2014).

[85] Y. Wu, Synthesis of $SrTiO_3$ nanoparticles for photocatalytic applications, Highlights in Science, Engineering and Technology **116**, 9 (2024).

[86] W. Jauch and A. Palmer, Anomalous zero-point motion in $SrTiO_3$: Results from γ-ray diffraction, Phys. Rev. B **60**, 2961 (1999).

[87] S. Keshri, L. Joshi, and S. S. Rajput, Studies on $La_{0.67}Ca_{0.33}MnO_3$–$SrTiO_3$ composites using two-phase model, J. Alloys Compd. **509**, 5796 (2011).

[88] S. A. Howard, J. K. Yau, and H. U. Anderson, Structural characteristics of $Sr_{1-x}La_xTi_{3+\delta}$ as a function of oxygen partial pressure at 1400 °C, J. Appl. Phys. **65**, 1492 (1989).

[89] Yu. A. Abramov, V. G. Tsirelson, V. E. Zavodnik, S. A. Ivanov, and Brown I. D., The chemical bond and atomic displacements in $SrTiO_3$ from X-ray diffraction analysis, Acta Crystallogr. B **51**, 942 (1995).

[90] J. M. Kiat and T. Roisnel, Rietveld analysis of strontium titanate in the Müller state, Journal of Physics: Condensed Matter **8**, 3471 (1996).

[91] L. Fang, W. Dong, F. Zheng, and M. Shen, Effects of Gd substitution on microstructures and low temperature dielectric relaxation behaviors of $SrTiO_3$ ceramics, J. Appl. Phys. **112**, (2012).

[92] J. Brous, I. Fankuchen, and E. Banks, Rare earth titanates with a perovskite structure, Acta Crystallogr. **6**, 67 (1953).

[93] M. Schmidbauer, A. Kwasniewski, and J. Schwarzkopf, High-precision absolute lattice parameter determination of $SrTiO_3$, $DyScO_3$ and $NdGaO_3$ single crystals, Acta Crystallogr. B **68**, 8 (2012).

[94] S. Parida, S. K. Rout, V. Subramanian, P. K. Barhai, N. Gupta, and V. R. Gupta, Structural, microwave dielectric properties and dielectric resonator antenna studies of $Sr(Zr_xTi_{1-x})O_3$ ceramics, J. Alloys Compd. **528**, 126 (2012).

[95] R. H. Mitchell, A. R. Chakhmouradian, and P. M. Woodward, Crystal chemistry of perovskite-type compounds in the tausonite-loparite series, $(Sr_{1-2x}Na_xLa_x)TiO_3$, Phys. Chem. Miner. **27**, 583 (2000).

[96] J. Hutton and R. J. Nelmes, High-resolution studies of cubic perovskites by elastic neutron diffraction. II. $SrTiO_3$, $KMnF_3$, $RbCaF_3$ and $CsPbCl_3$, Journal of Physics C: Solid State Physics **14**, 1713 (1981).

[97] K. Tsuda and M. Tanaka, Refinement of crystal structure parameters using convergent-beam electron diffraction: the low-temperature phase of $SrTiO_3$, Acta Crystallogr. A **51**, 7 (1995).

[98] Z. Yang, D. Lee, J. Yue, J. Gabel, T.-L. Lee, R. D. James, S. A. Chambers, and B. Jalan, Epitaxial $SrTiO_3$ films with dielectric constants exceeding 25,000, Proceedings of the National Academy of Sciences **119**, (2022).

[99] M. Ahtee and L. Unonius, The structure of $NaTaO_3$ by X-ray powder diffraction, Acta Crystallographica Section A **33**, 150 (1977).

[100] M. Ahtee and C. N. W. Darlington, Structures of $NaTaO_3$ by neutron powder diffraction, Acta Crystallogr. B **36**, 1007 (1980).

[101] S. W. Arulnesan, P. Kayser, B. J. Kennedy, and K. S. Knight, The impact of room temperature polymorphism in K doped $NaTaO_3$ on structural phase transition behaviour, J. Solid State Chem. **238**, 109 (2016).

[102] B. J. Kennedy, A. K. Prodjosantoso, and C. J. Howard, Powder neutron diffraction study of the high temperature phase transitions in $NaTaO_3$, Journal of Physics: Condensed Matter **11**, 6319 (1999).

[103] V. Shanker, S. L. Samal, G. K. Pradhan, C. Narayana, and A. K. Ganguli, Nanocrystalline $NaNbO_3$ and $NaTaO_3$: Rietveld studies, Raman spectroscopy and dielectric properties, Solid State Sci. **11**, 562 (2009).

[104] F. Brisse, D. J. Stewart, V. Seidl, and O. Knop, Pyrochlores. VIII. Studies of some 2–5 Pyrochlores and Related Compounds and Minerals, Can. J. Chem. **50**, 3648 (1972).

[105] K. Łukaszewicz, A. Pietraszko, J. Stepień-Damm, and N. N. Kolpakova, Temperature dependence of the crystal structure and dynamic disorder of cadmium in cadmium pyroniobates [$Cd_2Nb_2O_7$ and $Cd_2Ta_2O_7$], Mater. Res. Bull. **29**, 987 (1994).

[106] S. N. Al-Refaie and S. A. Alboon, On the Analysis of $Cd_2Nb_2O_7$ Dielectric Dispersion, Phys. Procedia **25**, 15 (2012).

[107] N. Takani and H. Yamane, Structure analysis of $CaTi_{1-x}Sn_xO_3$ ($x = 0.0$–$1.0$) solid solutions, Powder Diffr. **29**, 254 (2014).

[108] X. Liu and RobertC. Liebermann, X-ray powder diffraction study of $CaTiO_3$ perovskite at high temperatures, Phys. Chem. Miner. **20**, (1993).

[109] M. Yashima and R. Ali, Structural phase transition and octahedral tilting in the calcium titanate perovskite $CaTiO_3$, Solid State Ion. **180**, 120 (2009).

[110] A. Chandra and D. Pandey, Evolution of crystallographic phases in the system $(Pb_{1-x}Ca_x)TiO_3$: A Rietveld study, J. Mater. Res. **18**, 407 (2003).

[111] M. Tsubota, F. Iga, K. Uchihira, T. Nakano, S. Kura, T. Takabatake, S. Kodama, H. Nakao, and Y. Murakami, Coupling between Orbital and Lattice Degrees of Freedom in $Y_{1-x}Ca_xTiO_3$ ($0<x\leq0.75$): A Resonant X-ray Scattering Study, J. Physical Soc. Japan **74**, 3259 (2005).

[112] K. S. Knight, Structural and thermoelastic properties of $CaTiO_3$ perovskite between 7K and 400K, J. Alloys Compd. **509**, 6337 (2011).

[113] R. C. S. Costa, A. D. S. Bruno Costa, F. N. A. Freire, M. R. P. Santos, J. S. Almeida, R. S. T. M. Sohn, J. M. Sasaki, and A. S. B. Sombra, Structural properties of $CaTi_{1-x}(Nb_{2/3}Li_{2/3})_xO_{3-\delta}$ (CNLTO) and $CaTi_{1-x}(Nb_{1/2}Ln_{1/2})_xO_3$ (Ln=Fe (CNFTO), Bi (CNBTO)), modified dielectric ceramics for microwave applications, Physica B Condens. Matter **404**, 1409 (2009).

[114] S. Yoon et al., Improved photoluminescence and afterglow of $CaTiO_3:Pr^{3+}$ by ammonia treatment, Opt. Mater. Express **3**, 248 (2013).

[115] V. Vashook, L. Vasylechko, M. Knapp, H. Ullmann, and U. Guth, Lanthanum doped calcium titanates: synthesis, crystal structure, thermal expansion and transport properties, J. Alloys Compd. **354**, 13 (2003).

[116] R. Ali and M. Yashima, Space group and crystal structure of the Perovskite $CaTiO_3$ from 296 to 1720 K, J. Solid State Chem. **178**, 2867 (2005).

[117] K. S. Knight, PARAMETERIZATION OF THE CRYSTAL STRUCTURES OF CENTROSYMMETRIC ZONE-BOUNDARY-TILTED PEROVSKITES: AN ANALYSIS IN TERMS OF SYMMETRY-ADAPTED BASIS-VECTORS OF THE CUBIC ARISTOTYPE PHASE, The Canadian Mineralogist **47**, 381 (2009).

[118] A. R. Chakhmouradian and R. H. Mitchell, A Structural Study of the Perovskite Series $CaTi_{1-2x}Fe_xNb_xO_3$, J. Solid State Chem. **138**, 272 (1998).

[119] R. H. Buttner and E. N. Maslen, Electron difference density and structural parameters in $CaTiO_3$, Acta Crystallogr. B **48**, 644 (1992).

[120] V. Vashook, D. Nitsche, L. Vasylechko, J. Rebello, J. Zosel, and U. Guth, Solid state synthesis, structure and transport properties of compositions in the $CaRu_{1-x}Ti_xO_{3-\delta}$ system, J. Alloys Compd. **485**, 73 (2009).

[121] L. H. Oliveira, A. P. de Moura, T. M. Mazzo, M. A. Ramírez, L. S. Cavalcante, S. G. Antonio, W. Avansi, V. R. Mastelaro, E. Longo, and J. A. Varela, Structural refinement and photoluminescence properties of irregular cube-like $(Ca_{1-x}Cu_x)TiO_3$ microcrystals synthesized by the microwave–hydrothermal method, Mater. Chem. Phys. **136**, 130 (2012).

[122] H. Yang, Y. Ohishi, K. Kurosaki, H. Muta, and S. Yamanaka, Thermomechanical properties of calcium series perovskite-type oxides, J. Alloys Compd. **504**, 201 (2010).

[123] H. F. Kay and P. C. Bailey, Structure and properties of $CaTiO_3$, Acta Crystallogr. **10**, 219 (1957).

[124] H. Y. Zhou, X. Q. Liu, X. L. Zhu, and X. M. Chen, $CaTiO_3$ linear dielectric ceramics with greatly enhanced dielectric strength and energy storage density, Journal of the American Ceramic Society **101**, 1999 (2018).

[125] J. E. F. S. Rodrigues, P. J. Castro, P. S. Pizani, W. R. Correr, and A. C. Hernandes, Structural ordering and dielectric properties of $Ba_3CaNb_2O_9$-based microwave ceramics, Ceram. Int. **42**, 18087 (2016).

[126] J. E. Rodrigues, D. M. Bezerra, and A. C. Hernandes, Ordering effect on the electrical properties of stoichiometric $Ba_3CaNb_2O_9$-based perovskite ceramics, Ceram. Int. **43**, 14015 (2017).

[127] A. P. Gorshkov, N. S. Volkova, D. G. Fukina, S. B. Levichev, and L. A. Istomin, Impurity defect absorption and photochromic effect in $KNbWO_6$, J. Solid State Chem. **298**, 122099 (2021).

[128] S. Coh et al., Si-compatible candidates for high-κ dielectrics with the Pbnm perovskite structure, Phys. Rev. B **82**, 064101 (2010).

[129] N. A. Spaldin, A beginner's guide to the modern theory of polarization, J. Solid State Chem. **195**, 2 (2012).

[130] M. A. Ghebouli, B. Ghebouli, A. Bouhemadou, M. Fatmi, and K. Bouamama, Structural, electronic, optical and thermodynamic properties of $Sr_xCa_{1-x}O$, $Ba_xSr_{1-x}O$ and $Ba_xCa_{1-x}O$ alloys, J. Alloys Compd. **509**, 1440 (2011).

[131] I. Errea, M. Calandra, and F. Mauri, First-Principles Theory of Anharmonicity and the Inverse Isotope Effect in Superconducting Palladium-Hydride Compounds, Phys. Rev. Lett. **111**, 177002 (2013).

[132] M. Sharma, R. Resta, and R. Car, Intermolecular Dynamical Charge Fluctuations in Water: A Signature of the H-Bond Network, Phys. Rev. Lett. **95**, 187401 (2005).

[133] A. Togo, First-principles Phonon Calculations with Phonopy and Phono3py, J. Physical Soc. Japan 92, (2023).

[134] K. T. Schütt, F. Arbabzadah, S. Chmiela, K. R. Müller, and A. Tkatchenko, Quantum-chemical insights from deep tensor neural networks, Nat. Commun. **8**, 13890 (2017).

[135] M. Geiger and T. Smidt, e3nn: Euclidean Neural Networks, ArXiv Preprint ArXiv:2207.09453 (2022).

[136] T. Xie and J. C. Grossman, Crystal Graph Convolutional Neural Networks for an Accurate and Interpretable Prediction of Material Properties, Phys. Rev. Lett. **120**, 145301 (2018).

[137] S. Elfwing, E. Uchibe, and K. Doya, Sigmoid-weighted linear units for neural network function approximation in reinforcement learning, Neural Networks **107**, 3 (2018).

[138] D. P. Kingma and J. Ba, Adam: A Method for Stochastic Optimization, ArXiv Preprint ArXiv:1412.6980 (2017).

[139] P. E. Blöchl, Projector augmented-wave method, Phys. Rev. B **50**, 17953 (1994).

[140] G. Kresse and J. Furthmüller, Efficient iterative schemes for ab initio total-energy calculations using a plane-wave basis set, Phys. Rev. B **54**, 11169 (1996).

[141] C. Chen, W. Ye, Y. Zuo, C. Zheng, and S. P. Ong, Graph Networks as a Universal Machine Learning Framework for Molecules and Crystals, Chemistry of Materials **31**, 3564 (2019).

[142] D. C. Wallace, *Thermodynamics of Crystals*, Dover edition (Dover Publications, New York, 1998).

[143] J. M. Ziman, *Electrons and Phonons* (Oxford University Press, Oxford, U.K., 2001).

[144] S. H. Wemple, Polarization Fluctuations and the Optical-Absorption Edge in $BaTiO_3$, Phys. Rev. B **2**, 2679 (1970).

[145] M. Balog, M. Schieber, M. Michman, and S. Patai, The chemical vapour deposition and characterization of $ZrO_2$ films from organometallic compounds, Thin Solid Films **47**, 109 (1977).

[146] S. Maj, Energy gap and density in $SiO_2$ polymorphs, Phys. Chem. Miner. **15**, 271 (1988).

[147] M. Batzill and U. Diebold, The surface and materials science of tin oxide, Prog. Surf. Sci. **79**, 47 (2005).

[148] Y. Hinuma, Y. Kumagai, I. Tanaka, and F. Oba, Effects of composition, crystal structure, and surface orientation on band alignment of divalent metal oxides: A first-principles study, Phys. Rev. Mater. **2**, 124603 (2018).

[149] Y. Hinuma, T. Gake, and F. Oba, Band alignment at surfaces and heterointerfaces of $Al_2O_3$, $Ga_2O_3$, $In_2O_3$, and related group-III oxide polymorphs: A first-principles study, Phys. Rev. Mater. **3**, 084605 (2019).

[150] A. Merchant, S. Batzner, S. S. Schoenholz, M. Aykol, G. Cheon, and E. D. Cubuk, Scaling deep learning for materials discovery, Nature **624**, 80 (2023).

**Physics-Based Factorized Machine Learning for Predicting Ionic Dielectric Tensors**

Atsushi Takigawa[1,2,4], Shin Kiyohara[1,4]* and Yu Kumagai[1,3]**

[1]Institute for Materials Research, Tohoku University, 2-1-1 Katahira, Aoba-ku, Sendai 980-8577, Japan
[2]Graduate School of Engineering, Tohoku University, 6-6 Aramaki Aoba, Aoba-ku, Sendai 980-8579, Japan
[3]Organization for Advanced Studies, Tohoku University, 2-1-1 Katahira, Aoba-ku, Sendai 980-8577, Japan
[4]These authors contributed equally: Atsushi Takigawa and Shin Kiyohara
* sin@tohoku.ac.jp
**yukumagai@tohoku.ac.jp

# Supplemental Figures and Tables

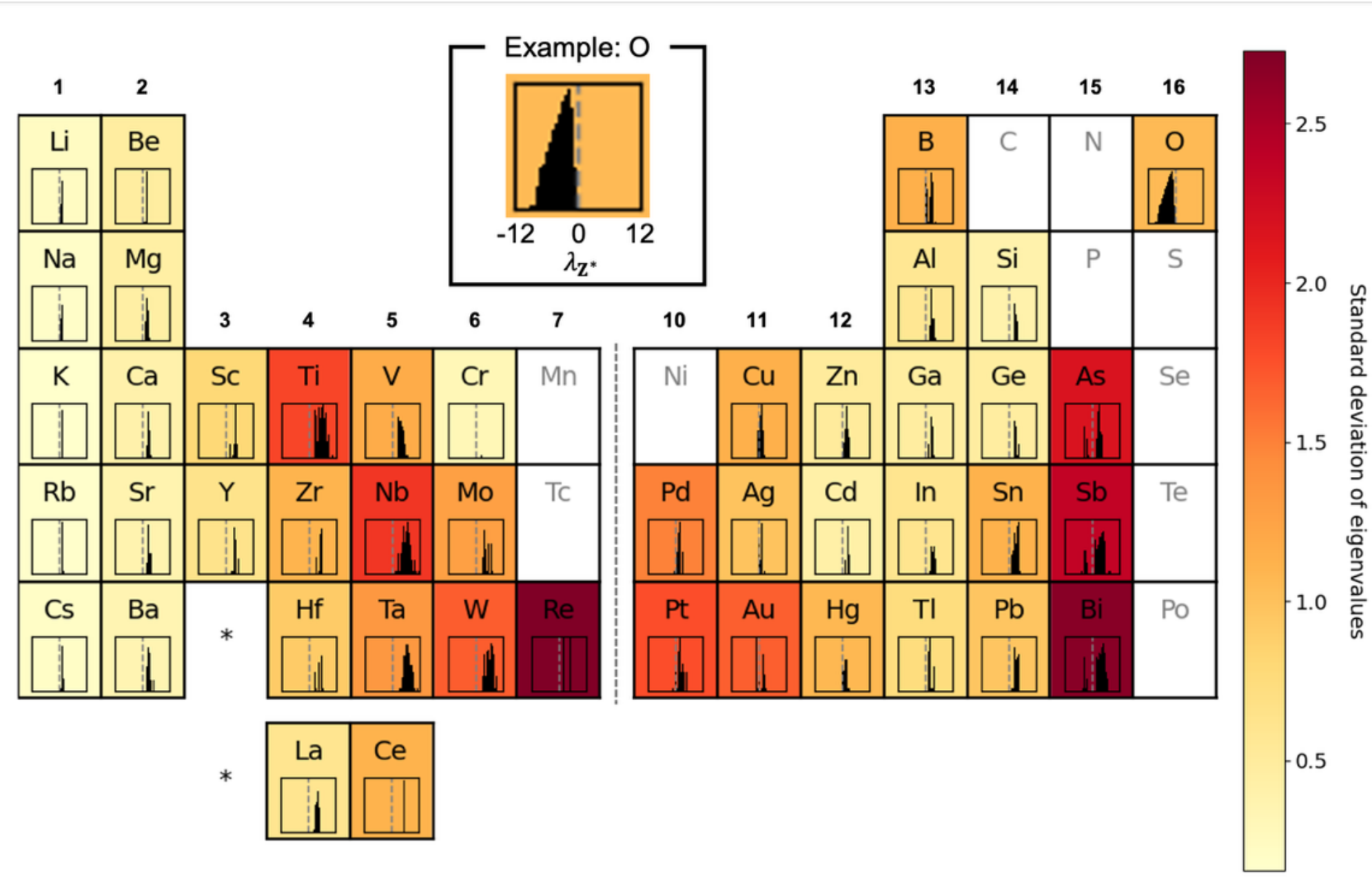

Supplemental FIG. 1. Histograms showing the distributions of Born effective charge (BEC) eigenvalues for the 928 oxides in the training dataset, grouped by element. The horizontal axis of each histogram spans from −12 to 12, with the dotted line indicating zero (see oxygen as an example). The color of each tile represents the standard deviation of the eigenvalues for the corresponding element, indicating the importance of accurate ML prediction.

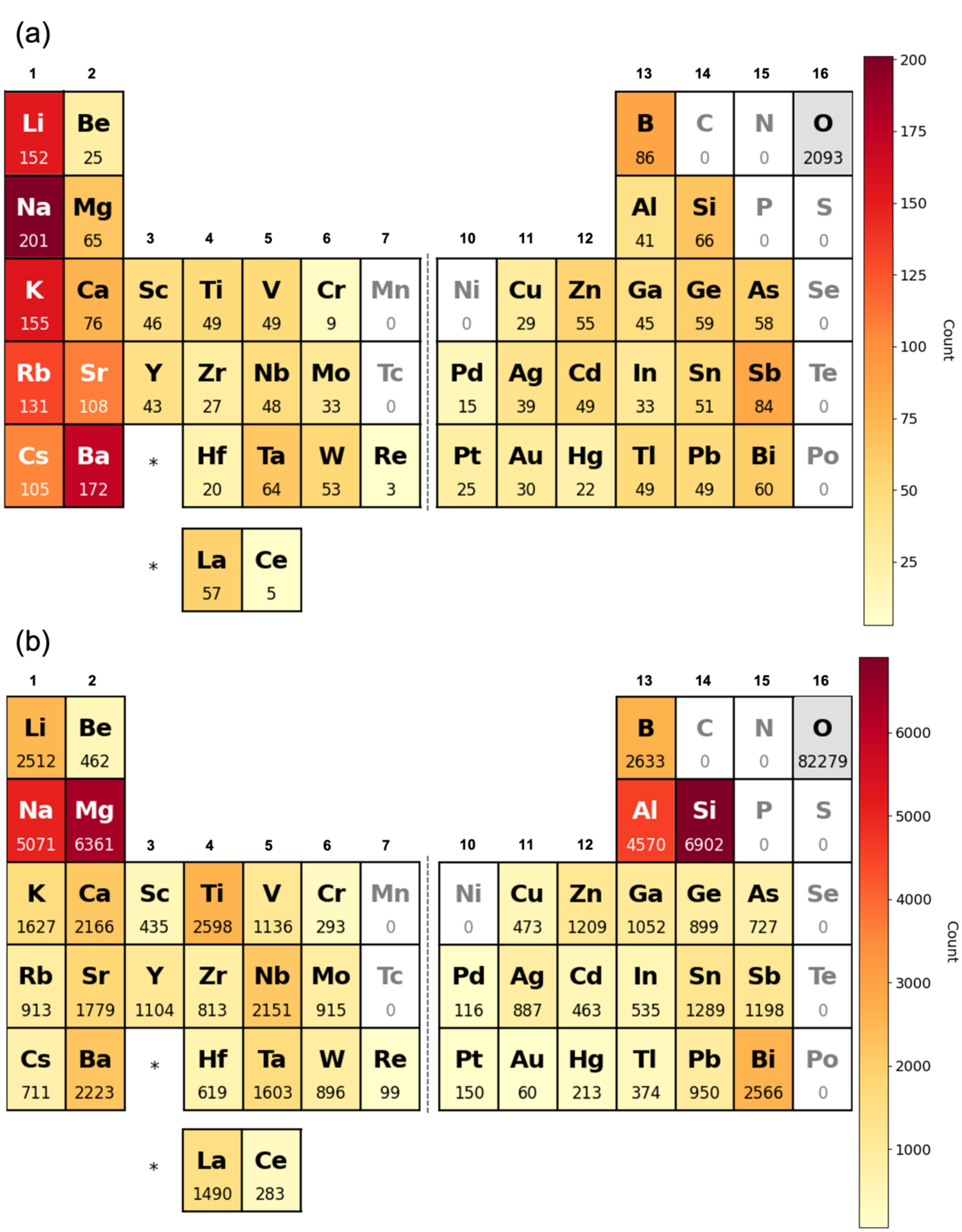

Supplemental FIG. 2. Heat maps showing the numbers of unique atomic sites for each element appearing in (a) the 928 oxides in the training dataset and (b) the 8,717 oxides in the screening dataset. Oxygen is displayed in grey because its number is exceptionally high.

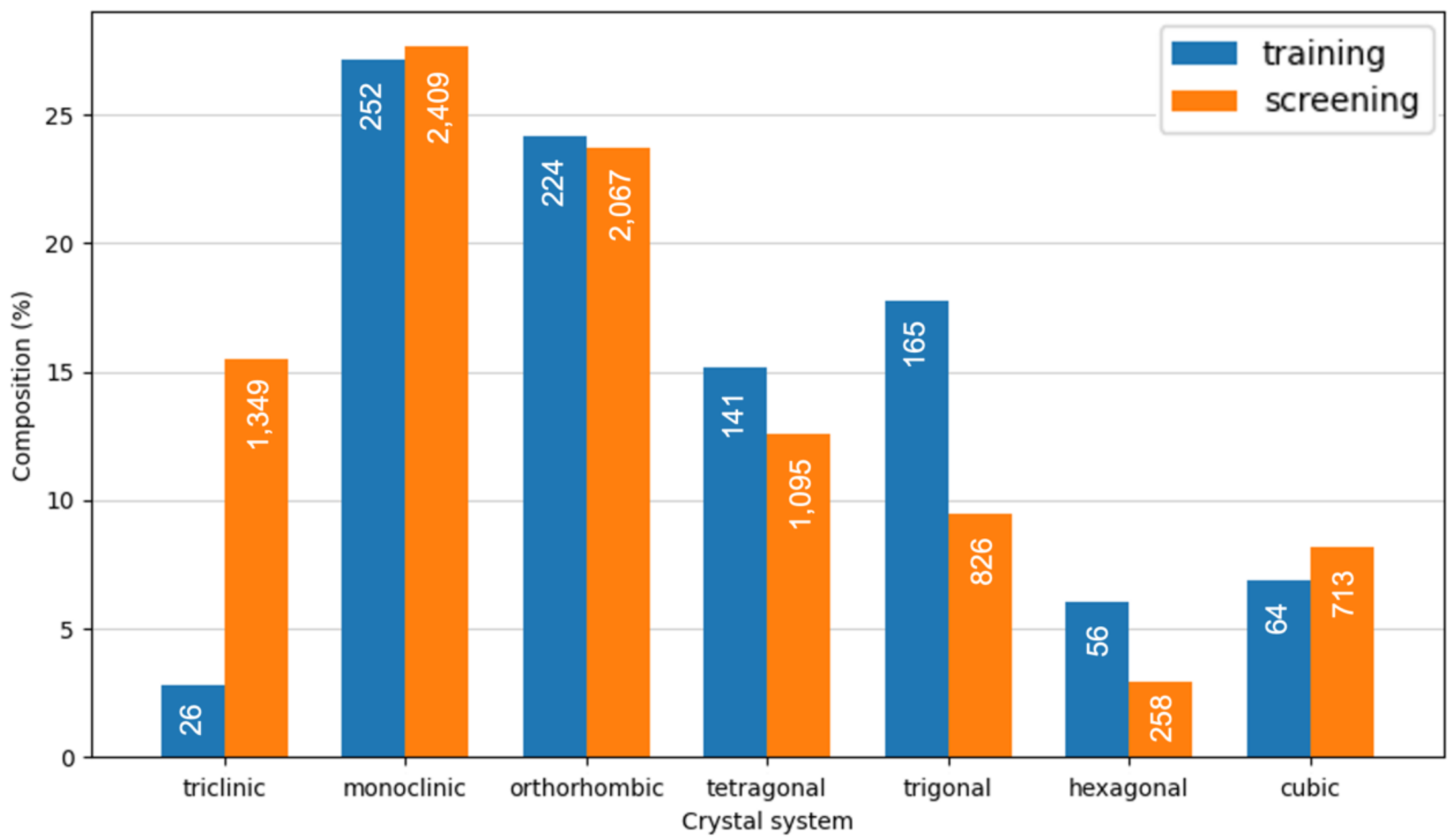

Supplemental FIG. 3. Bar charts showing the composition of crystal systems in the training dataset (928 entries; blue bars) and the dataset used for virtual screening (8,717 entries; orange bars). The numbers shown above each bar indicate the number of entries belonging to the corresponding crystal system.

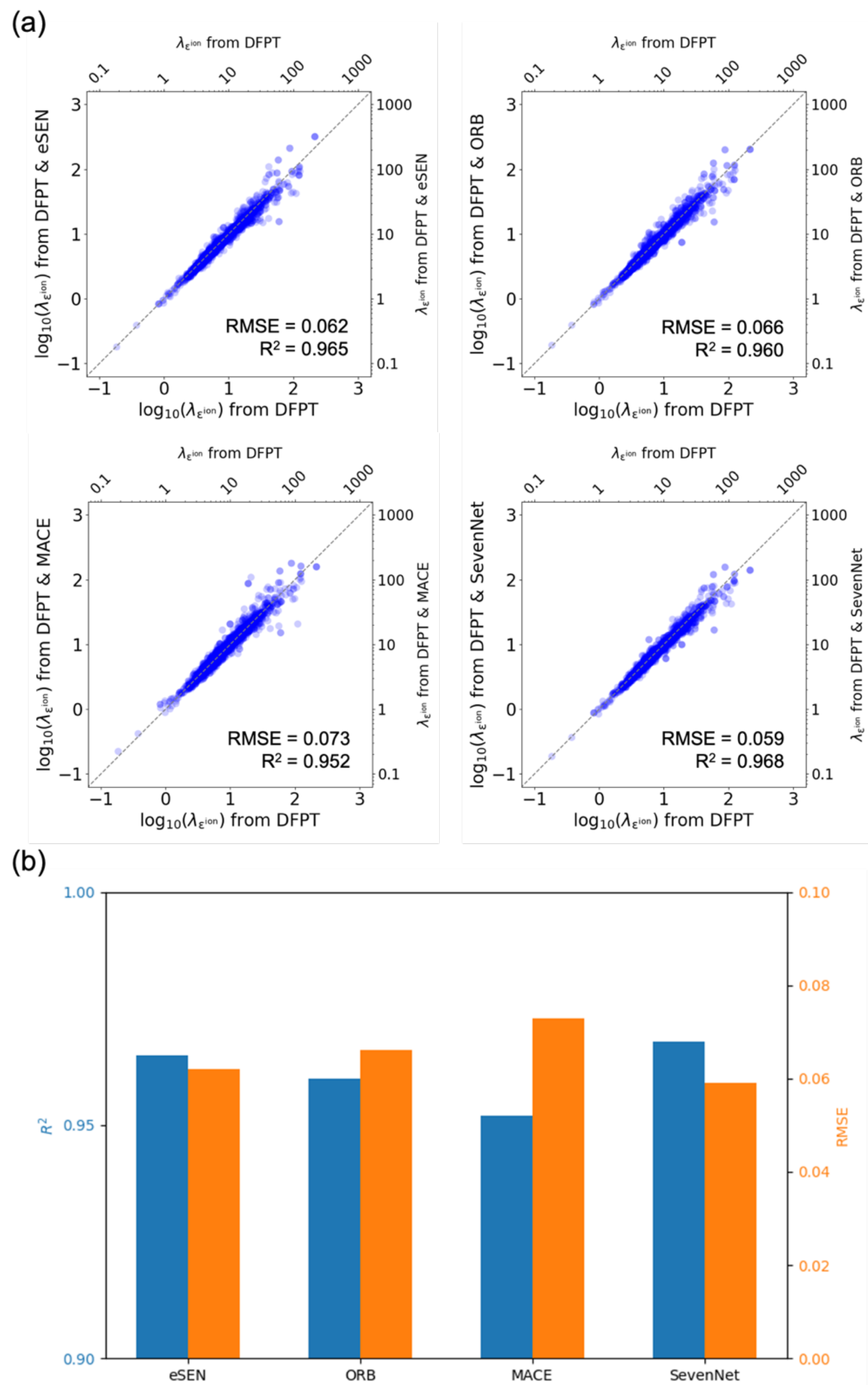

Supplemental FIG. 4. Comparison of MLPs for predicting the ionic dielectric tensors. Oxides predicted to be dynamically unstable by at least one MLP were excluded, leaving 786 oxides for evaluation. The ionic dielectric tensor was predicted using Eq. (2) in the main text. BECs obtained from first-principles calculations were used to isolate the errors from the phonon properties (force constant matrices). (a) Parity plots for eigenvalues of ionic dielectric tensors. (b) The blue and orange bar graph heights represent the average $R^2$ and RMSE, respectively.

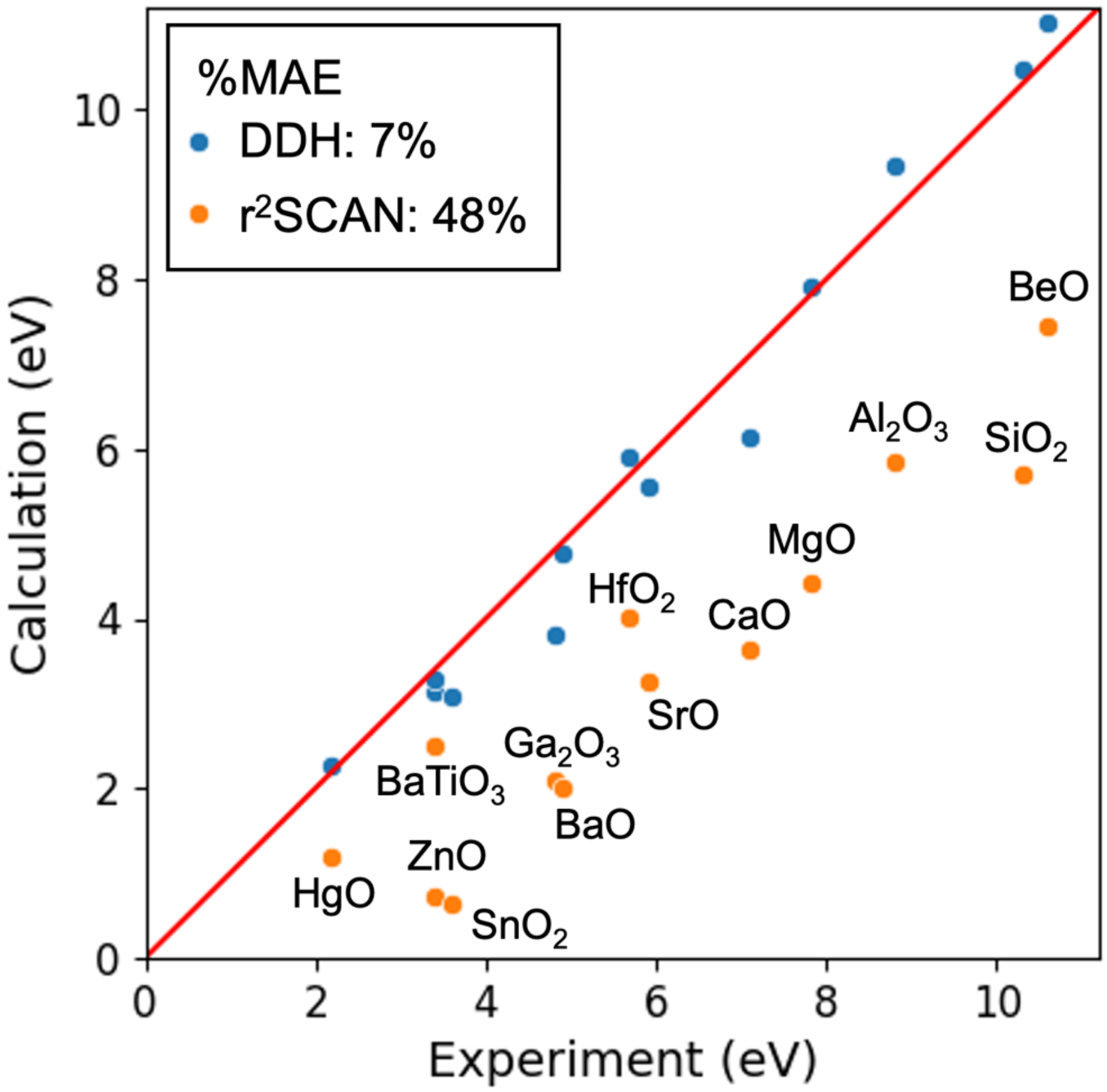

Supplemental FIG. 5. Band gap values computed using r$^2$SCAN, and dielectric-dependent hybrid (DDH) functionals compared with experimental measurements [1–6]. The values calculated with r$^2$SCAN were obtained from the Materials Project Database [7]. The corresponding mean absolute percentage errors (%MAE) are also presented.

# Supplemental Notes

## Supplemental Note 1: Prediction accuracies for dielectric constants

The prediction accuracies of the BECs, electronic dielectric constant, ionic dielectric constant, and total dielectric constant were evaluated for both tensorial quantities and scalar values, using 5-fold cross-validation. The initial learning rates optimized for each task are summarized in Supplemental Table 2.

The parity plots for the EGNN predictions of the electronic dielectric constants are shown in Supplemental Fig. 6. Those for the direct model, for both electronic and ionic dielectric constants, are shown in Supplemental Figs. 7 and 8, while the parity plots for the joint model of the ionic dielectric constant are presented in Supplemental Fig. 9. Finally, in Supplemental Fig. 10, the parity plots of the total dielectric constant predicted by summing the electronic dielectric constant (from direct model) and the ionic dielectric constant (from joint model),. In all the cases, RMSE and $R^2$ are shown.

Supplemental TABLE 1. Initial learning rates optimized for each task.

| **Task** | BEC | Electronic dielectric constant | Ionic dielectric constant |
|---|---|---|---|
| Initial learning rate | $4\times10^{-2}$ | $4\times10^{-2}$ | $2\times10^{-2}$ |

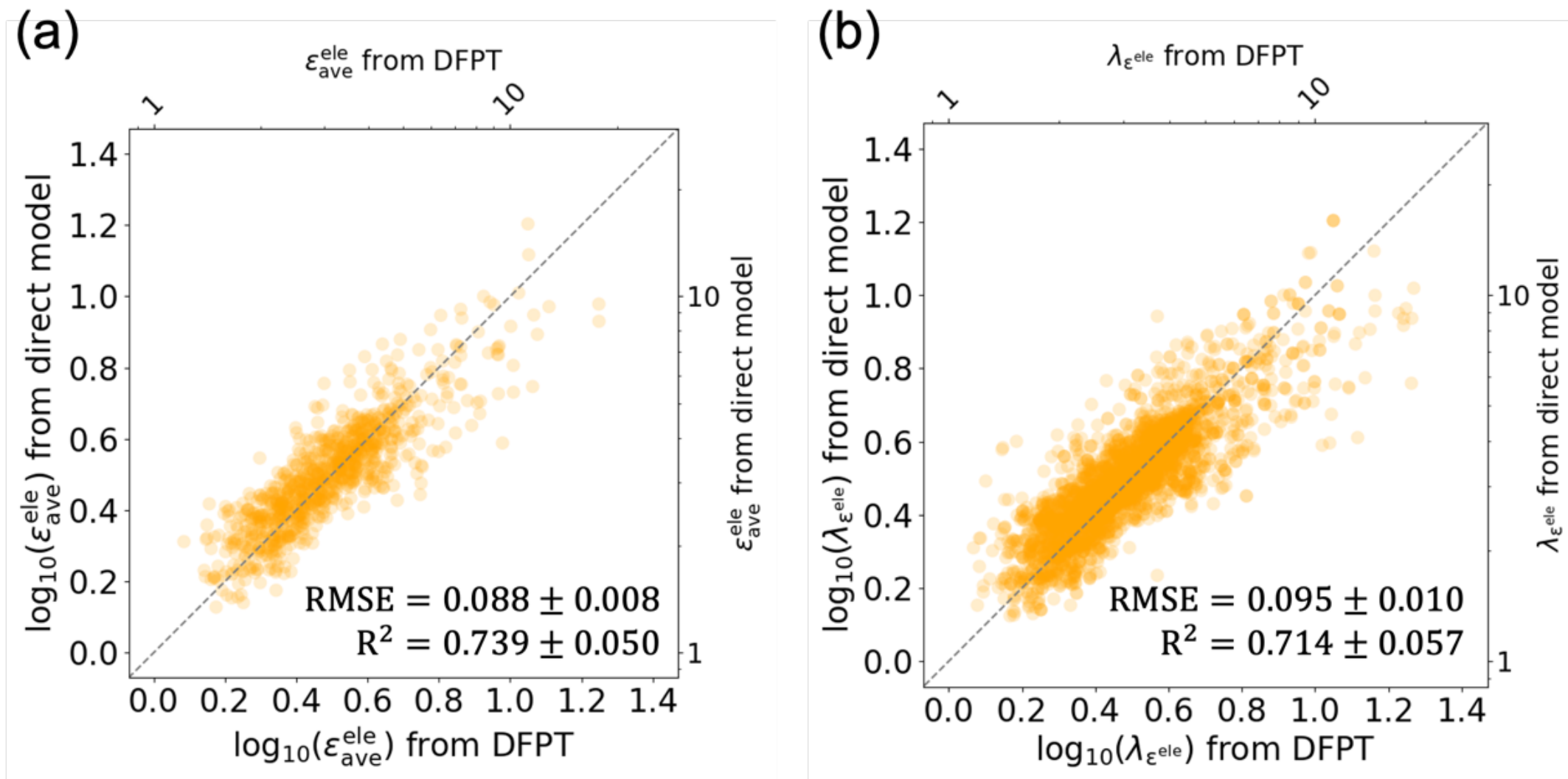

Supplemental FIG. 6. Parity plots of electronic dielectric constants. The prediction accuracies of the electronic dielectric constants using the EGNN are evaluated for (a) scalar values and (b) eigenvalues.

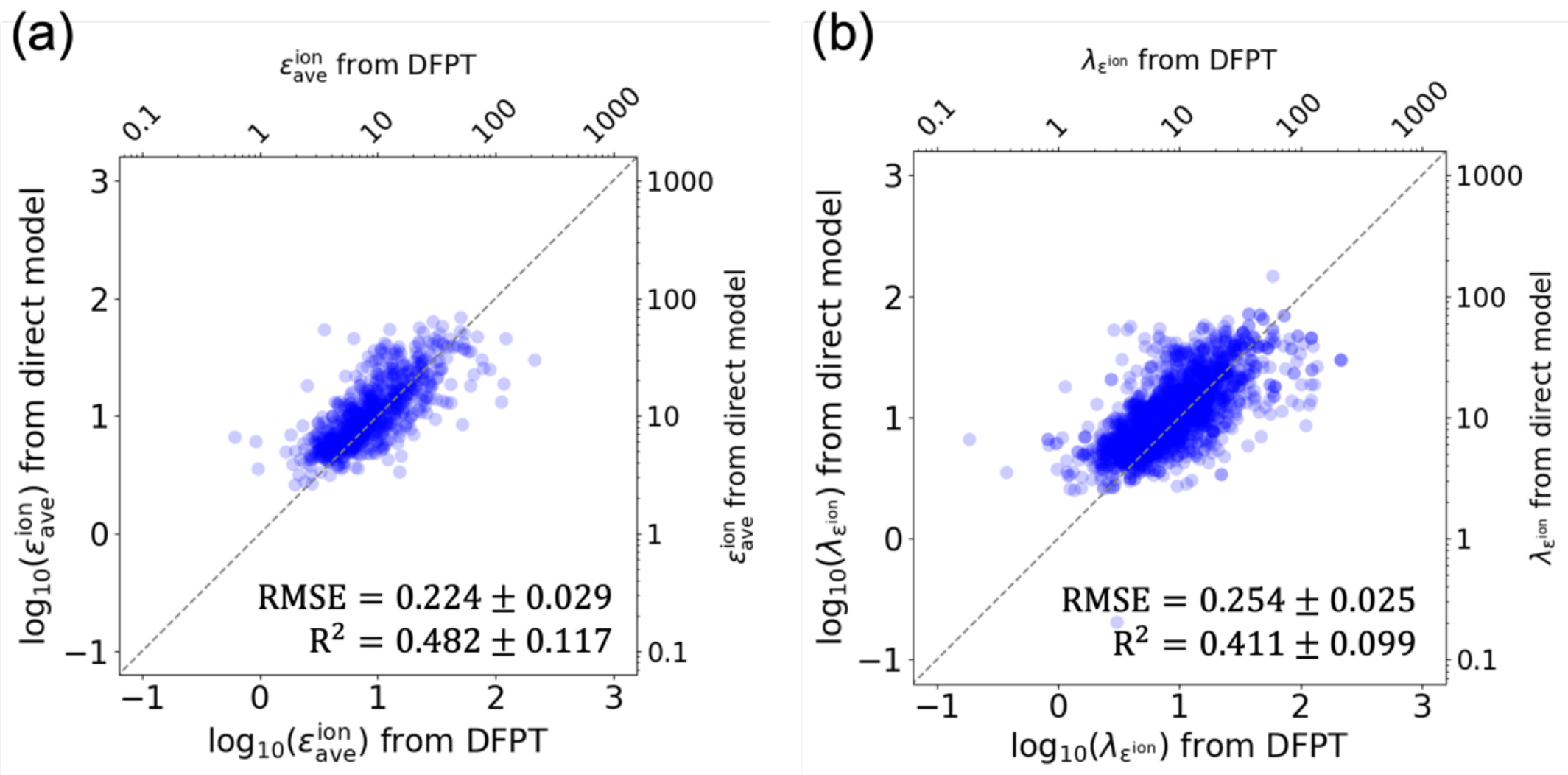

Supplemental FIG. 7. Parity plots of ionic dielectric constants (direct model). The prediction accuracies of the ionic dielectric constants using the constructed EGNN (direct model) are evaluated for (a) scalar values and (b) eigenvalues. The prediction accuracies were evaluated for the 839 oxides that SevenNet predicted to be stable.

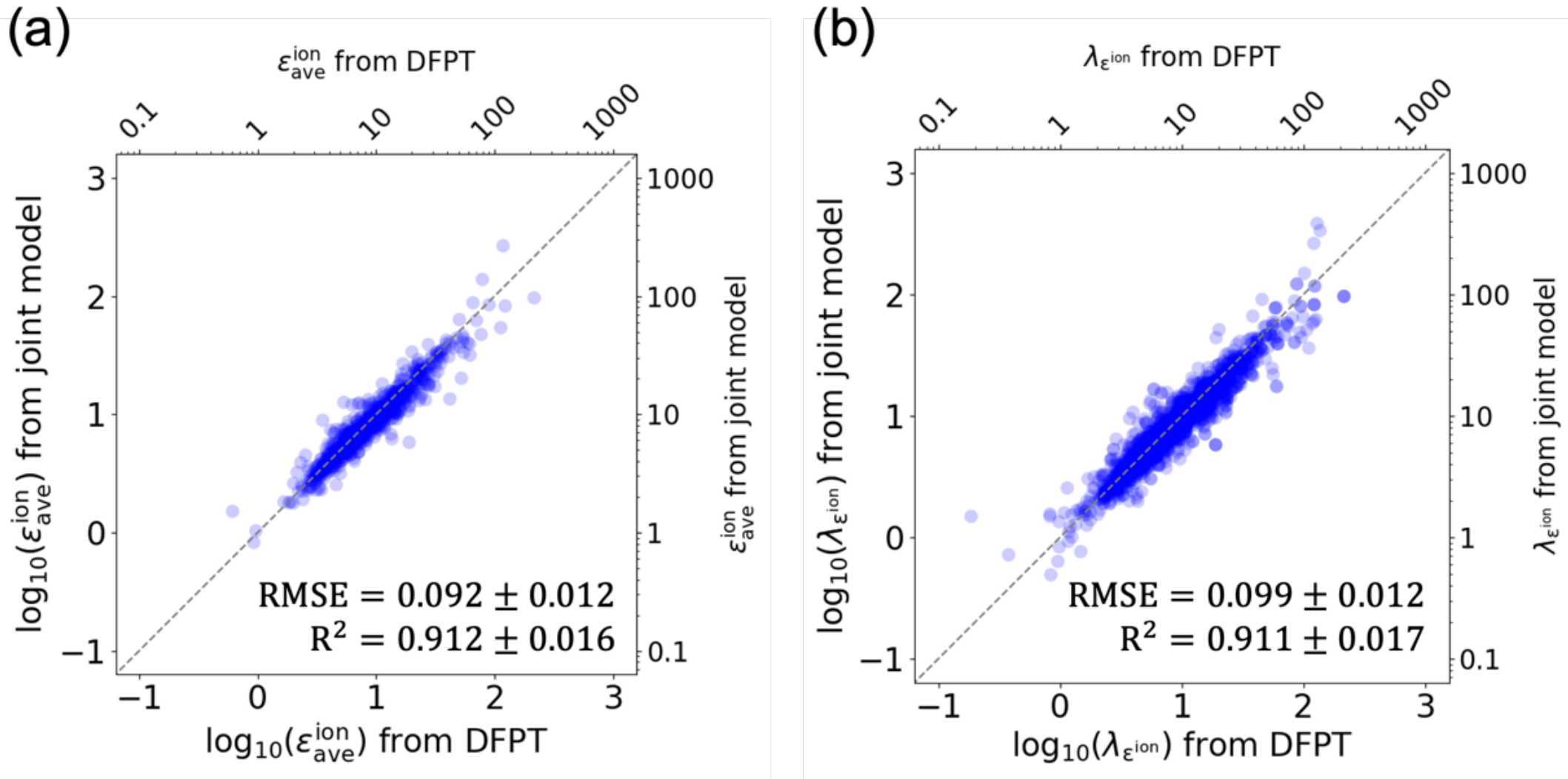

Supplemental FIG. 8. Parity plots of ionic dielectric constants (joint model). The prediction accuracies of the ionic dielectric constants using the constructed joint model are evaluated for (a) scalar values and (b) eigenvalues. The prediction accuracies were evaluated for the 839 oxides that SevenNet predicted to be stable.

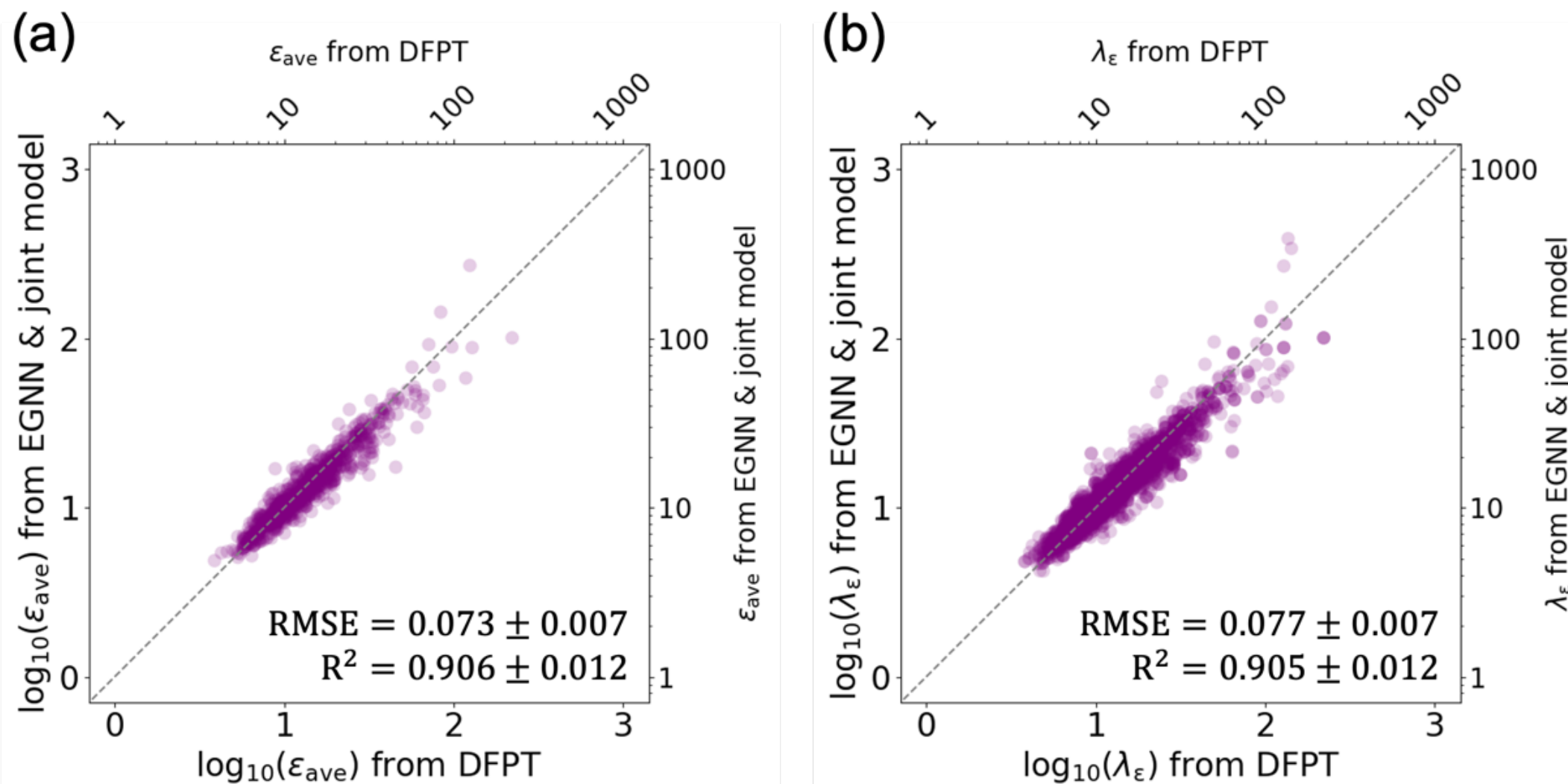

Supplemental FIG. 9. Parity plots of total dielectric constants. The prediction accuracies of the total dielectric constants using the model for the electronic dielectric constants and the joint model for the ionic dielectric constants are evaluated for (a) scalar values and (b) eigenvalues. The prediction accuracies were evaluated for the 839 oxides that SevenNet predicted to be stable.

## Supplemental Note 2: Contributions of phonons and Born effective charges to the errors in the prediction of ionic contributions to the dielectric constants

As described in Sec. IIA, we decomposed the prediction errors into two contributions: (1) combinations of ML-predicted BECs with DFPT-calculated phonons, and (2) combinations of DFPT-calculated BECs with ML-predicted phonons. The parity plots are shown in Supplemental Fig. 10, and corresponding MAE, RMSE, and $R^2$ metrics are summarized in Supplemental Table 2. See the main text for detailed discussion.

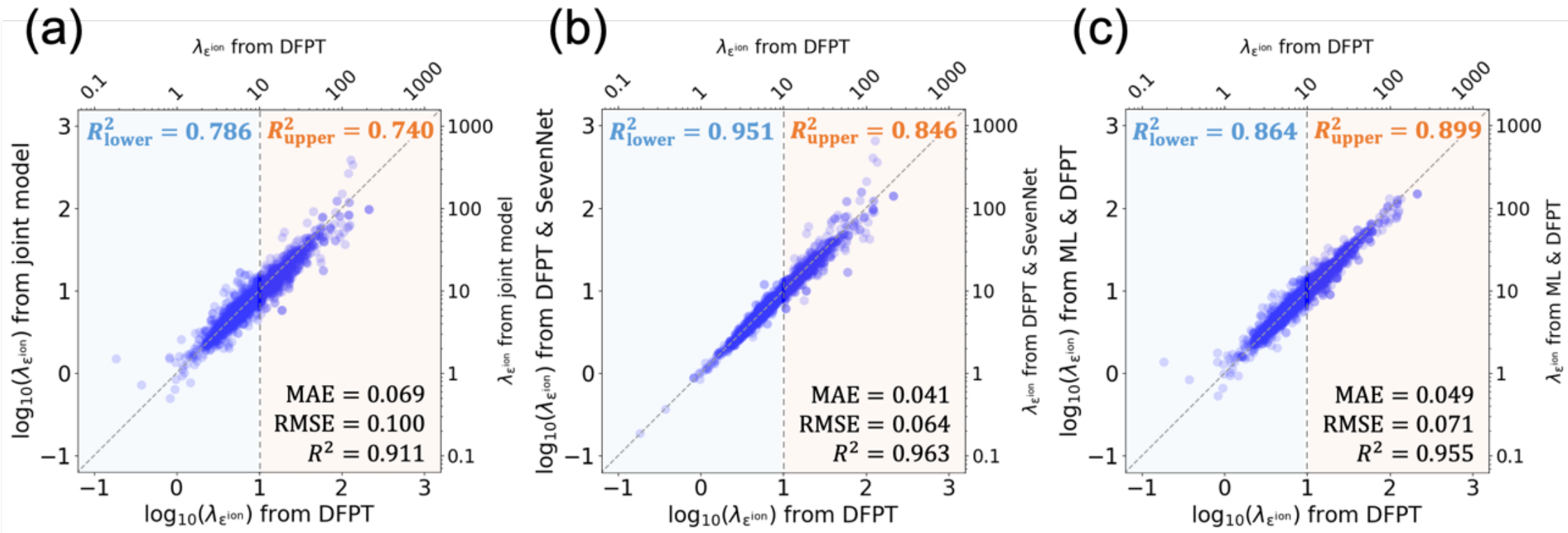

FIG. 10. Parity plots for the eigenvalues of the ionic contribution to the dielectric tensor, $\boldsymbol{\varepsilon}^{ion}$, obtained from all test sets of 5-fold cross-validation used for BEC training. Only the 839 oxides predicted to be dynamically stable by SevenNet are considered. In all panels, $\boldsymbol{\varepsilon}^{ion}$ is diagonalized and the base-10 logarithm of each eigenvalue is plotted. (a) Both phonon properties and BECs are predicted by machine-learning (ML) models. (b) Phonon properties are predicted by ML, while BECs are obtained from DFPT. (c) Phonon frequencies are obtained from DFPT, while BECs are predicted by ML. The vertical dashed line at $\lambda_{\boldsymbol{\varepsilon}^{ion}} = 10$ separates the lower and upper eigenvalue regimes used for region-wise evaluation. The upper-eigenvalue regime exhibits noticeably reduced predictive accuracy, highlighting the increased modeling difficulty for large $\boldsymbol{\varepsilon}^{ion}$ contributions.

Supplemental TABLE 2. Decomposition of $\boldsymbol{\varepsilon}^{ion}$ prediction errors. MAE, RMSE, and R² for the predicted eigenvalues of the dielectric tensor ($\lambda_{\boldsymbol{\varepsilon}^{ion}}$) are reported. See the caption in Table 2 in the main text.

| Dataset | Metrics | Total | Phonon | BEC |
|---|---|---|---|---|
| All | MAE | 0.069 | 0.041 | 0.049 |
| | RMSE | 0.100 | 0.064 | 0.071 |
| | $R^2$ | 0.911 | 0.963 | 0.955 |
| High-$\boldsymbol{\varepsilon}^{ion}$ | MAE | 0.086 | 0.060 | 0.054 |
| | RMSE | 0.120 | 0.092 | 0.075 |
| | $R^2$ | 0.740 | 0.846 | 0.899 |
| Low-$\boldsymbol{\varepsilon}^{ion}$ | MAE | 0.059 | 0.031 | 0.047 |
| | RMSE | 0.086 | 0.041 | 0.069 |
| | $R^2$ | 0.786 | 0.951 | 0.864 |

## Supplemental Note 3: Detailed information of the 31 identified high-κ oxides and the nine oxides in the training dataset.

31 identified high-κ oxides by the virtual screening and 9 oxides in the training dataset with $\mathrm{FoM}_{\mathrm{DDH}}^{\mathrm{DFPT}} \geq 350$ are listed in Supplemental Table 3. The "MPD-id" denotes the identifier in the Materials Project Database (MPD) [7]. "GNoME" indicates oxides included in the Graph Networks for Materials Exploration Database [8]. "Δion" represents the fraction of the total dielectric constant attributable to the ionic contribution. "Space Group" and "$E_{\mathrm{hull}}$" correspond to the values reported in the MPD.

Supplemental TABLE 3. Complete list of 31 identified oxides.

| **Formula** | $SrZrO_3$ | $Ba_3TiNb_4O_{15}$ | $Pb_2TiZrO_6$ | $Sr_3Hf_2O_7$ | $NaNbO_3$ | $AgTaO_3$ | $CdTiO_3$ | $Ta_2Zn_4O_9$ |
|---|---|---|---|---|---|---|---|---|
| **MPD-id** | mp-1020622 | mp-557072 | mp-1079416 | mp-779517 | mp-3671 | mp-1187189 | mp-20940 | mp-778795 |
| **Space group** | *P4/mbm* | *P4bm* | *P4mm* | *I4/mmm* | *Pbcm* | $Pm\bar{3}m$ | $Pmc2_1$ | $P12_1/c1$ |
| $\boldsymbol{E}_{\mathbf{hull}}$ | 0.02 | 0.02 | 0.04 | 0.02 | 0.00 | 0.05 | 0.04 | 0.10 |
| $\boldsymbol{E}_{\mathbf{g}}^{\mathbf{DDH}}$ | 5.1 | 3.7 | 2.8 | 5.6 | 3.7 | 3.1 | 3.4 | 4.7 |
| $\boldsymbol{\varepsilon}_{\mathbf{ave}}^{\mathbf{ele,DFPT}}$ | 3.7 | 5.0 | 5.8 | 3.2 | 4.8 | 5.7 | 6.7 | 4.2 |
| $\boldsymbol{\varepsilon}_{\mathbf{ave}}^{\mathbf{ion,DFPT}}$ | 535 | 516 | 547 | 263 | 377 | 382 | 290 | 190 |
| $\boldsymbol{\varepsilon}_{\mathbf{ave}}^{\mathbf{DFPT}}$ | 539 | 522 | 554 | 267 | 383 | 389 | 297 | 196 |
| **Δion** | 99.1% | 98.8% | 98.8% | 98.4% | 98.5% | 98.3% | 97.4% | 97.3% |
| $\mathbf{FoM}_{\mathbf{DDH}}^{\mathbf{DFPT}}$ | 2752 | 1912 | 1540 | 1490 | 1421 | 1200 | 1001 | 911 |
| **GNoME** | | | | | | | | |
| **Synthesized** | | ✓ [9] | | | ✓ [10–13] | | ✓ [14] | |
| $\boldsymbol{\varepsilon}_{\mathbf{ave}}^{\mathbf{Exp}}$ | | 900 [15–19] | | | 200 [20] | | | |
| **Structure** | Perovskite | $ReO_3$ (tungsten bronze) | Perovskite | Layered-perovskite | Perovskite-derived | Perovskite | Perovskite | |

| **Formula** | $Sr_3CaO_4$ | $Al_2O_3$ | CdTiO3 | $Ta_4Cd_3HgO_{14}$ | $WO_3$ | $SrZrO_3$ | $CaTiO_3$ | $Pb_3Ti_2ZrO_9$ |
|---|---|---|---|---|---|---|---|---|
| **MPD-id** | mp-1079940 | mp-642363 | mp-5052 | mp-3211899 | mp-2383052 | mp-1068742 | mp-556003 | mp-1215328 |
| **Space group** | $Pm\bar{3}m$ | $Cmcm$ | $Pna2_1$ | $R\bar{3}m$ | $Cmcm$ | $P4mm$ | $Imma$ | $C1m1$ |
| $\boldsymbol{E}_{\mathbf{hull}}$ | 0.03 | 0.28 | 0.04 | -- | 0.00 | 0.05 | 0.01 | 0.03 |
| $\boldsymbol{E}_{\mathbf{g}}^{\mathbf{DDH}}$ | 5.4 | 6.9 | 3.2 | 3.0 | 1.6 | 5.1 | 3.2 | 3.5 |
| $\boldsymbol{\varepsilon}_{\mathbf{ave}}^{\mathbf{ele}}$ | 2.9 | 2.5 | 6.7 | 5.0 | 5.3 | 3.6 | 5.6 | 6.0 |
| $\boldsymbol{\varepsilon}_{\mathbf{ave}}^{\mathbf{ion}}$ | 134 | 98 | 186 | 170 | 326 | 96 | 153 | 132 |
| $\boldsymbol{\varepsilon}_{\mathbf{ave}}$ | 138 | 102 | 193 | 176 | 332 | 101 | 159 | 139 |
| **Δion** | 97.2% | 96.5% | 96.0% | 96.6% | 98.1% | 95.5% | 95.8% | 94.9% |
| $\mathbf{FoM}_{\mathbf{DDH}}^{\mathbf{DFPT}}$ | 747 | 699 | 611 | 522 | 517 | 515 | 509 | 487 |
| **GNoME** | | | | ✓ | | | | |
| **Synth.** | | ✓ [21] | ✓ [22–24] | | | | | |
| $\boldsymbol{\varepsilon}_{\mathbf{ave}}^{\mathbf{Exp}}$ | | | 2100 [25] | | | | | |
| **Structure** | Caswellsilverite-like | $CaIrO_3$ | Perovskite | | $ReO_3$ | Perovskite-derived | Perovskite-derived | Perovskite-derived |

| **Formula** | $TiO_2$ | $Pb_5Ti_3Zr_2O_{15}$ | $Pb_2TiZrO_6$ | $Pb_2TiZrO_6$ | $Hf_3Ti_2(PbO_3)_5$ | NaNbO3 | $SrTa_4Cd_3O_{14}$ | $Ta_2CdPbO_7$ |
|---|---|---|---|---|---|---|---|---|
| **MPD-id** | mp-1008677 | mp-1216039 | mp-1215309 | mp-1079056 | mp-1224603 | mp-1188528 | mp-3210467 | mp-3206064 |
| **Space group** | $Fm\bar{3}m$ | $Pmm2$ | $P1$ | $P4mm$ | $C1m1$ | $Pna2_1$ | $R\bar{3}m$ | $Imma$ |
| $\boldsymbol{E}_{\mathbf{hull}}$ | 0.31 | 0.02 | 0.03 | 0.02 | 0.02 | 0.00 | -- | -- |
| $\boldsymbol{E}_{\mathbf{g}}^{\mathbf{DDH}}$ | 1.5 | 3.2 | 3.7 | 3.4 | 3.6 | 3.6 | 3.9 | 3.4 |
| $\boldsymbol{\varepsilon}_{\mathbf{ave}}^{\mathbf{ele}}$ | 9.9 | 5.7 | 5.6 | 5.6 | 5.5 | 4.9 | 4.5 | 5.2 |
| $\boldsymbol{\varepsilon}_{\mathbf{ave}}^{\mathbf{ion}}$ | 302 | 144 | 124 | 129 | 119 | 115 | 106 | 120 |
| $\boldsymbol{\varepsilon}_{\mathbf{ave}}$ | 313 | 151 | 130 | 135 | 125 | 121 | 112 | 126 |
| **Δion** | 96.5% | 95.5% | 95.0% | 95.1% | 94.8% | 95.1% | 95.1% | 95.1% |
| $\mathbf{FoM}_{\mathbf{DDH}}^{\mathbf{DFPT}}$ | 481 | 477 | 476 | 465 | 457 | 439 | 430 | 427 |
| **GNoME** | | | | | | | ✓ | ✓ |
| **Synth.** | | | | | | ✓ [10] | | |
| $\boldsymbol{\varepsilon}_{\mathbf{ave}}^{\mathbf{Exp}}$ | | | | | | | | |
| **Structure** | Fluorite | Perovskite-derived | Perovskite-derived | Perovskite-derived | Perovskite-derived | Perovskite | | |

| **Formula** | $Y_2Ti_2O_7$ | $Ta_2HgPbO_7$ | $CaTa_4Cd_3O_{14}$ | $LiBi_4Nb_3O_{14}$ | $SrHfO_3$ | $NaCaTaTiO_6$ | $LaYO_3$ |
|---|---|---|---|---|---|---|---|
| **MPD-id** | mp-5373 | mp-3205975 | mp-3216088 | mp-559052 | mp-550908 | mp-39712 | mp-10429 |
| **Space group** | $Fd\bar{3}m1$ | *Imma* | $R\bar{3}m$ | *C12/c1* | *P4/mbm* | *P1c1* | *Pnma* |
| $\boldsymbol{E}_{\mathbf{hull}}$ | 0.01 | -- | -- | -- | 0.01 | 0.01 | 0.04 |
| $\boldsymbol{E}_{\mathbf{g}}^{\mathbf{DDH}}$ | 4.0 | 3.0 | 4.1 | 3.7 | 5.6 | 3.7 | 6.5 |
| $\boldsymbol{\varepsilon}_{\mathbf{ave}}^{\mathbf{ele}}$ | 5.2 | 5.5 | 4.5 | 5.8 | 3.4 | 4.9 | 3.6 |
| $\boldsymbol{\varepsilon}_{\mathbf{ave}}^{\mathbf{ion}}$ | 99 | 121 | 86 | 94 | 61 | 91 | 49 |
| $\boldsymbol{\varepsilon}_{\mathbf{ave}}$ | 105 | 127 | 92 | 101 | 65 | 97 | 54 |
| **Δion** | 94.1% | 94.8% | 94.0% | 93.3% | 93.3% | 93.9% | 91.4% |
| $\mathbf{FoM}_{\mathbf{DDH}}^{\mathbf{DFPT}}$ | 416 | 384 | 372 | 371 | 367 | 353 | 351 |
| **GNoME** | | ✓ | ✓ | | | | |
| **Synth.** | ✓ [26–32] | | | ✓ [33] | | | ✓ [34] |
| $\boldsymbol{\varepsilon}_{\mathbf{ave}}^{\mathbf{Exp}}$ | 54 [35] | | | | | | |
| **Structure** | Pyrochlore | | | Layered-perovskite (Aurivillius-derived) | Perovskite-derived | Perovskite-derived | Perovskite |

Supplemental TABLE 4. Complete list of 9 oxides from the training dataset.

| **Formula** | $KTaO_3$ | $SrTiO_3$ | $NaTaO_3$ | $Cd_2Ta_2O_7$ | $Ba_3Nb_2CdO_9$ | $CaTiO_3$ | $Ba_3CaNb_2O_9$ | $KNbWO_6$ | $TaTlWO_6$ |
|---|---|---|---|---|---|---|---|---|---|
| **MPD-id** | mp-3614 | mp-4651 | mp-3858 | mp-5548 | mp-1214502 | mp-4019 | mp-1214569 | mp-1223396 | mp-1217830 |
| **Space group** | $Pm\bar{3}m$ | $I4/mcm$ | $Pnma$ | $Fd\bar{3}m1$ | $P\bar{3}m1$ | $Pnma$ | $P\bar{3}m1$ | $Ima2$ | $Ima2$ |
| $\boldsymbol{E_{\mathrm{hull}}}$ | -- | -- | -- | -- | -- | -- | -- | -- | -- |
| $\boldsymbol{E_{\mathrm{g}}^{\mathrm{DDH}}}$ | 3.5 | 3.1 | 4.3 | 3.6 | 4.5 | 3.5 | 4.4 | 4.7 | 4.6 |
| $\boldsymbol{\varepsilon_{\mathrm{ave}}^{\mathrm{ele}}}$ | 4.5 | 5.7 | 4.3 | 4.8 | 4.2 | 5.6 | 4.0 | 4.0 | 4.4 |
| $\boldsymbol{\varepsilon_{\mathrm{ave}}^{\mathrm{ion}}}$ | 216 | 213 | 113 | 123 | 96 | 118 | 92 | 79 | 77 |
| $\boldsymbol{\varepsilon_{\mathrm{ave}}}$ | 221 | 220 | 118 | 129 | 101 | 125 | 97 | 84 | 82 |
| **Δion** | 97.5% | 97.0% | 95.5% | 95.5% | 94.9% | 94.7% | 94.9% | 94.1% | 93.4% |
| $\mathbf{FoM_{DDH}^{DFPT}}$ | 785 | 678 | 512 | 466 | 457 | 433 | 431 | 392 | 376 |
| **Synth.** | ✓ [36,37] | ✓ [38–50] | ✓ [51–54] | ✓ [55,56] | | ✓ [23,57–73] | ✓ [74] | ✓ [75] | |
| $\boldsymbol{\varepsilon_{\mathrm{ave}}^{\mathrm{Exp}}}$ | 5000 [76] | 4500 [77] | 147 [78] | 1200 [79] | | 180 [80] | 40.3 [81] | | |
| **Structure** | Perovskite | Perovskite | Perovskite | Pyrochlore | Perovskite-derived | Perovskite | Perovskite-derived | Pyrochlore-derived | |

## Supplemental References

[1] S. H. Wemple, Polarization Fluctuations and the Optical-Absorption Edge in $BaTiO_3$, Phys. Rev. B **2**, 2679 (1970).

[2] M. Balog, M. Schieber, M. Michman, and S. Patai, The chemical vapour deposition and characterization of $ZrO_2$ films from organometallic compounds, Thin Solid Films **47**, 109 (1977).

[3] S. Maj, Energy gap and density in $SiO_2$ polymorphs, Phys. Chem. Miner. **15**, 271 (1988).

[4] M. Batzill and U. Diebold, The surface and materials science of tin oxide, Prog. Surf. Sci. **79**, 47 (2005).

[5] Y. Hinuma, Y. Kumagai, I. Tanaka, and F. Oba, Effects of composition, crystal structure, and surface orientation on band alignment of divalent metal oxides: A first-principles study, Phys. Rev. Mater. **2**, 124603 (2018).

[6] Y. Hinuma, T. Gake, and F. Oba, Band alignment at surfaces and heterointerfaces of $Al_2O_3$, $Ga_2O_3$, $In_2O_3$, and related group-III oxide polymorphs: A first-principles study, Phys. Rev. Mater. **3**, 084605 (2019).

[7] A. Jain et al., Commentary: The Materials Project: A materials genome approach to accelerating materials innovation, APL Mater. **1**, (2013).

[8] A. Merchant, S. Batzner, S. S. Schoenholz, M. Aykol, G. Cheon, and E. D. Cubuk, Scaling deep learning for materials discovery, Nature **624**, 80 (2023).

[9] P. B. Jamieson and S. C. Abrahams, Re-examination of the crystal structure of ferroelectric tetragonal bronze-type $Ba_6Ti_2Nb_8O_{30}$, Acta Crystallogr. B **24**, 984 (1968).

[10] K. E. Johnston, C. C. Tang, J. E. Parker, K. S. Knight, P. Lightfoot, and S. E. Ashbrook, The Polar Phase of $NaNbO_3$: A Combined Study by Powder Diffraction, Solid-State NMR, and First-Principles Calculations, J. Am. Chem. Soc. **132**, 8732 (2010).

[11] H. Xu, Y. Su, M. Lou Balmer, and A. Navrotsky, A New Series of Oxygen-Deficient Perovskites in the $NaTi_xNb_{1-x}O_{3-0.5x}$ System: Synthesis, Crystal Chemistry, and Energetics, Chemistry of Materials **15**, 1872 (2003).

[12] A. C. Sakowski-Cowley, K. Łukaszewicz, and H. D. Megaw, The structure of sodium niobate at room temperature, and the problem of reliability in pseudosymmetric structures, Acta Crystallogr. B **25**, 851 (1969).

[13] S. K. Mishra, N. Choudhury, S. L. Chaplot, P. S. R. Krishna, and R. Mittal, Competing antiferroelectric and ferroelectric interactions in $NaNbO_3$: Neutron diffraction and theoretical studies, Phys. Rev. B **76**, 024110 (2007).

[14] Y. J. Shan, H. Mori, R. Wang, W. Luan, H. Imoto, Mitsuruitoh, and T. Nakamura, Powder X-ray diffraction study of ferroelectric phase transition in perovskite oxide $CdTiO_3$, Ferroelectrics **259**, 85 (2001).

[15] R. Someya, Y. J. Shan, and K. Tezuka, Crystal structure and dielectric properties of $Ba_{6-x}Ca_xTi_2Nb_8O_{30}$ tungsten bronze ceramics, Journal of the Ceramic Society of Japan **131**, 22170 (2023).

[16] Y. Yuan, X. M. Chen, and Y. J. Wu, Diffused ferroelectrics of $Ba_6Ti_2Nb_8O_{30}$ and $Sr_6Ti_2Nb_8O_{30}$ with filled tungsten-bronze structure, J. Appl. Phys. **98**, (2005).

[17] G. L. Roberts, R. J. Cava, W. F. Peck, and J. J. Krajewski, Dielectric properties of barium titanium niobates, J. Mater. Res. **12**, 526 (1997).

[18] Y. Itoh, S. Miyazawa, T. Yamada, and H. Iwasaki, Dielectric Property of $Ba_6Ti_2Nb_8O_{30}$ Single Crystal, Jpn. J. Appl. Phys. **9**, 157 (1970).

[19] K. Praveena and K. B. R. Varma, Ferroelectric and optical properties of $Ba_5Li_2Ti_2Nb_8O_{30}$ ceramics potential for memory applications, Journal of Materials Science: Materials in Electronics **25**, 3103 (2014).

[20] D. O. Mishchuk, O. I. V'yunov, O. V. Ovchar, and A. G. Belous, Structural and dielectric properties of solid solutions of sodium niobate in lanthanum and neodymium niobates, Inorganic Materials **40**, 1324 (2004).

[21] R. Caracas and R. E. Cohen, Prediction of a new phase transition in $Al_2O_3$ at high pressures, Geophys. Res. Lett. **32**, 1 (2005).

[22] B. J. Kennedy, Q. Zhou, and M. Avdeev, The ferroelectric phase of $CdTiO_3$: A powder neutron diffraction study, J. Solid State Chem. **184**, 2987 (2011).

[23] S. Sasaki, C. T. Prewitt, J. D. Bass, and W. A. Schulze, Orthorhombic perovskite $CaTiO_3$ and $CdTiO_3$: structure and space group, Acta Crystallogr. C **43**, 1668 (1987).

[24] H. F. Kay and J. L. Miles, The structure of cadmium titanate and sodium tantalate, Acta Crystallogr. **10**, 213 (1957).

[25] T. Sugai and M. Wada, Single Crystal Growth and Some Properties of $Cd_2Ti_2O_5F_2$ and $CdTiO_3$, Jpn. J. Appl. Phys. **18**, 1709 (1979).

[26] A. Garbout, N. Kallel- Kchaou, and M. Férid, Relationship between the structural characteristics and photoluminescent properties of $LnEuTi_2O_7$ (Ln=Gd and Y) pyrochlores, J. Lumin. **169**, 359 (2016).

[27] L. Kong, Z. Zhang, M. de los Reyes, I. Karatchevtseva, G. R. Lumpkin, G. Triani, and R. D. Aughterson, Soft chemical synthesis and structural characterization of $Y_2Hf_xTi_{2-x}O_7$, Ceram. Int. **41**, 5309 (2015).

[28] F. Matteucci, G. Cruciani, M. Dondi, G. Baldi, and A. Barzanti, Crystal structural and optical properties of Cr-doped $Y_2Ti_2O_7$ and $Y_2Sn_2O_7$ pyrochlores, Acta Mater. **55**, 2229 (2007).

[29] O. Knop, F. Brisse, and L. Castelliz, Pyrochlores. V. Thermoanalytic, X-ray, neutron, infrared, and dielectric studies of $A_2Ti_2O_7$ titanates, Can. J. Chem. **47**, 971 (1969).

[30] B. J. Sun, Q. L. Liu, J. K. Liang, J. B. Li, L. N. Ji, J. Y. Zhang, Y. H. Liu, and G. H. Rao, Subsolidus phase relations in the ternary system $SnO_2$–$TiO_2$–$Y_2O_3$, J. Alloys Compd. **455**, 265 (2008).

[31] M. Glerup, O. F. Nielsen, and F. W. Poulsen, The Structural Transformation from the Pyrochlore Structure, $A_2B_2O_7$, to the Fluorite Structure, $AO_2$, Studied by Raman Spectroscopy and Defect Chemistry Modeling, J. Solid State Chem. **160**, 25 (2001).

[32] J. M. Farmer, L. A. Boatner, B. C. Chakoumakos, M.-H. Du, M. J. Lance, C. J. Rawn, and J. C. Bryan, Structural and crystal chemical properties of rare-earth titanate pyrochlores, J. Alloys Compd. **605**, 63 (2014).

[33] B. Muktha, M. H. Priya, G. Madras, and T. N. Guru Row, Synthesis, Structure, and Photocatalysis in a New Structural Variant of the Aurivillius Phase: $LiBi_4M_3O_{14}$ (M = Nb, Ta), J. Phys. Chem. B **109**, 11442 (2005).

[34] M. Bharathy, A. H. Fox, S. J. Mugavero, and H.-C. zur Loye, Crystal growth of inter-lanthanide $LaLn'O_3$ (Ln′=Y, Ho–Lu) perovskites from hydroxide fluxes, Solid State Sci. **11**, 651 (2009).

[35] T. Ting-ting, W. Li-xi, and Z. Qi-tu, Study on the composite and properties of $Y_2O_3$–$TiO_2$ microwave dielectric ceramics, J. Alloys Compd. **486**, 606 (2009).

[36] E. A. Zhurova, Y. Ivanov, V. Zavodnik, and V. Tsirelson, Electron density and atomic displacements in $KTaO_3$, Acta Crystallogr. B **56**, 594 (2000).

[37] A. Tkach, P. M. Vilarinho, and A. Almeida, Role of initial potassium excess on the properties of potassium tantalate ceramics, J. Eur. Ceram. Soc. **31**, 2303 (2011).

[38] Y. Wu, Synthesis of $SrTiO_3$ nanoparticles for photocatalytic applications, Highlights in Science, Engineering and Technology **116**, 9 (2024).

[39] W. Jauch and A. Palmer, Anomalous zero-point motion in $SrTiO_3$: Results from γ-ray diffraction, Phys. Rev. B **60**, 2961 (1999).

[40] S. Keshri, L. Joshi, and S. S. Rajput, Studies on $La_{0.67}Ca_{0.33}MnO_3$–$SrTiO_3$ composites using two-phase model, J. Alloys Compd. **509**, 5796 (2011).

[41] S. A. Howard, J. K. Yau, and H. U. Anderson, Structural characteristics of $Sr_{1-x}La_xTi_{3+\delta}$ as a function of oxygen partial pressure at 1400 °C, J. Appl. Phys. **65**, 1492 (1989).

[42] Yu. A. Abramov, V. G. Tsirelson, V. E. Zavodnik, S. A. Ivanov, and Brown I. D., The chemical bond and atomic displacements in $SrTiO_3$ from X-ray diffraction analysis, Acta Crystallogr. B **51**, 942 (1995).

[43] J. M. Kiat and T. Roisnel, Rietveld analysis of strontium titanate in the Müller state, Journal of Physics: Condensed Matter **8**, 3471 (1996).

[44] L. Fang, W. Dong, F. Zheng, and M. Shen, Effects of Gd substitution on microstructures and low temperature dielectric relaxation behaviors of $SrTiO_3$ ceramics, J. Appl. Phys. **112**, (2012).

[45] J. Brous, I. Fankuchen, and E. Banks, Rare earth titanates with a perovskite structure, Acta Crystallogr. **6**, 67 (1953).

[46] M. Schmidbauer, A. Kwasniewski, and J. Schwarzkopf, High-precision absolute lattice parameter determination of $SrTiO_3$, $DyScO_3$ and $NdGaO_3$ single crystals, Acta Crystallogr. B **68**, 8 (2012).

[47] S. Parida, S. K. Rout, V. Subramanian, P. K. Barhai, N. Gupta, and V. R. Gupta, Structural, microwave dielectric properties and dielectric resonator antenna studies of $Sr(Zr_xTi_{1-x})O_3$ ceramics, J. Alloys Compd. **528**, 126 (2012).

[48] R. H. Mitchell, A. R. Chakhmouradian, and P. M. Woodward, Crystal chemistry of perovskite-type compounds in the tausonite-loparite series, $(Sr_{1-2x}Na_xLa_x)TiO_3$, Phys. Chem. Miner. **27**, 583 (2000).

[49] J. Hutton and R. J. Nelmes, High-resolution studies of cubic perovskites by elastic neutron diffraction. II. $SrTiO_3$, $KMnF_3$, $RbCaF_3$ and $CsPbCl_3$, Journal of Physics C: Solid State Physics **14**, 1713 (1981).

[50] K. Tsuda and M. Tanaka, Refinement of crystal structure parameters using convergent-beam electron diffraction: the low-temperature phase of $SrTiO_3$, Acta Crystallogr. A **51**, 7 (1995).

[51] M. Ahtee and C. N. W. Darlington, Structures of $NaTaO_3$ by neutron powder diffraction, Acta Crystallogr. B **36**, 1007 (1980).

[52] S. W. Arulnesan, P. Kayser, B. J. Kennedy, and K. S. Knight, The impact of room temperature polymorphism in K doped $NaTaO_3$ on structural phase transition behaviour, J. Solid State Chem. **238**, 109 (2016).

[53] M. Ahtee and L. Unonius, The structure of $NaTaO_3$ by X-ray powder diffraction, Acta Crystallographica Section A **33**, 150 (1977).

[54] B. J. Kennedy, A. K. Prodjosantoso, and C. J. Howard, Powder neutron diffraction study of the high temperature phase transitions in $NaTaO_3$, Journal of Physics: Condensed Matter **11**, 6319 (1999).

[55] F. Brisse, D. J. Stewart, V. Seidl, and O. Knop, Pyrochlores. VIII. Studies of some 2–5 Pyrochlores and Related Compounds and Minerals, Can. J. Chem. **50**, 3648 (1972).

[56] K. Łukaszewicz, A. Pietraszko, J. Stepień-Damm, and N. N. Kolpakova, Temperature dependence of the crystal structure and dynamic disorder of cadmium in cadmium pyroniobates [$Cd_2Nb_2O_7$ and $Cd_2Ta_2O_7$], Mater. Res. Bull. **29**, 987 (1994).

[57] N. Takani and H. Yamane, Structure analysis of $CaTi_{1-x}Sn_xO_3$ (x = 0.0–1.0) solid solutions, Powder Diffr. **29**, 254 (2014).

[58] X. Liu and RobertC. Liebermann, X-ray powder diffraction study of $CaTiO_3$ perovskite at high temperatures, Phys. Chem. Miner. **20**, (1993).

[59] M. Yashima and R. Ali, Structural phase transition and octahedral tilting in the calcium titanate perovskite $CaTiO_3$, Solid State Ion. **180**, 120 (2009).

[60] A. Chandra and D. Pandey, Evolution of crystallographic phases in the system $(Pb_{1-x}Ca_x)TiO_3$: A Rietveld study, J. Mater. Res. **18**, 407 (2003).

[61] M. Tsubota, F. Iga, K. Uchihira, T. Nakano, S. Kura, T. Takabatake, S. Kodama, H. Nakao, and Y. Murakami, Coupling between Orbital and Lattice Degrees of Freedom in $Y_{1-x}Ca_xTiO_3$ ($0<x\leq0.75$): A Resonant X-ray Scattering Study, J. Physical Soc. Japan **74**, 3259 (2005).

[62] K. S. Knight, Structural and thermoelastic properties of $CaTiO_3$ perovskite between 7K and 400K, J. Alloys Compd. **509**, 6337 (2011).

[63] R. C. S. Costa, A. D. S. Bruno Costa, F. N. A. Freire, M. R. P. Santos, J. S. Almeida, R. S. T. M. Sohn, J. M. Sasaki, and A. S. B. Sombra, Structural properties of $CaTi_{1-x}(Nb_{2/3}Li_{2/3})xO_{3-\delta}$ (CNLTO) and $CaTi_{1-x}(Nb_{1/2}Ln_{1/2})_xO_3$ (Ln=Fe (CNFTO), Bi (CNBTO)), modified dielectric ceramics for microwave applications, Physica B Condens. Matter **404**, 1409 (2009).

[64] S. Yoon et al., Improved photoluminescence and afterglow of $CaTiO_3:Pr^{3+}$ by ammonia treatment, Opt. Mater. Express **3**, 248 (2013).

[65] V. Vashook, L. Vasylechko, M. Knapp, H. Ullmann, and U. Guth, Lanthanum doped calcium titanates: synthesis, crystal structure, thermal expansion and transport properties, J. Alloys Compd. **354**, 13 (2003).

[66] R. Ali and M. Yashima, Space group and crystal structure of the Perovskite $CaTiO_3$ from 296 to 1720 K, J. Solid State Chem. **178**, 2867 (2005).

[67] K. S. Knight, PARAMETERIZATION OF THE CRYSTAL STRUCTURES OF CENTROSYMMETRIC ZONE-BOUNDARY-TILTED PEROVSKITES: AN ANALYSIS IN TERMS OF SYMMETRY-ADAPTED BASIS-VECTORS OF THE CUBIC ARISTOTYPE PHASE, The Canadian Mineralogist **47**, 381 (2009).

[68] A. R. Chakhmouradian and R. H. Mitchell, A Structural Study of the Perovskite Series $CaTi_{1-2x}Fe_xNb_xO_3$, J. Solid State Chem. **138**, 272 (1998).

[69] R. H. Buttner and E. N. Maslen, Electron difference density and structural parameters in $CaTiO_3$, Acta Crystallogr. B **48**, 644 (1992).

[70] V. Vashook, D. Nitsche, L. Vasylechko, J. Rebello, J. Zosel, and U. Guth, Solid state synthesis, structure and transport properties of compositions in the $CaRu_{1-x}Ti_xO_{3-\delta}$ system, J. Alloys Compd. **485**, 73 (2009).

[71] L. H. Oliveira, A. P. de Moura, T. M. Mazzo, M. A. Ramírez, L. S. Cavalcante, S. G. Antonio, W. Avansi, V. R. Mastelaro, E. Longo, and J. A. Varela, Structural refinement and photoluminescence properties of irregular cube-like $(Ca_{1-x}Cu_x)TiO_3$ microcrystals synthesized by the microwave–hydrothermal method, Mater. Chem. Phys. **136**, 130 (2012).

[72] H. Yang, Y. Ohishi, K. Kurosaki, H. Muta, and S. Yamanaka, Thermomechanical properties of calcium series perovskite-type oxides, J. Alloys Compd. **504**, 201 (2010).

[73] H. F. Kay and P. C. Bailey, Structure and properties of $CaTiO_3$, Acta Crystallogr. **10**, 219 (1957).

[74] J. E. F. S. Rodrigues, P. J. Castro, P. S. Pizani, W. R. Correr, and A. C. Hernandes, Structural ordering and dielectric properties of $Ba_3CaNb_2O_9$-based microwave ceramics, Ceram. Int. **42**, 18087 (2016).

[75] A. P. Gorshkov, N. S. Volkova, D. G. Fukina, S. B. Levichev, and L. A. Istomin, Impurity defect absorption and photochromic effect in $KNbWO_6$, J. Solid State Chem. **298**, 122099 (2021).

[76] O. Aktas, S. Crossley, M. A. Carpenter, and E. K. H. Salje, Polar correlations and defect-induced ferroelectricity in cryogenic $KTaO_3$, Phys. Rev. B **90**, 165309 (2014).

[77] Z. Yang, D. Lee, J. Yue, J. Gabel, T.-L. Lee, R. D. James, S. A. Chambers, and B. Jalan, Epitaxial $SrTiO_3$ films with dielectric constants exceeding 25,000, Proceedings of the National Academy of Sciences **119**, (2022).

[78] V. Shanker, S. L. Samal, G. K. Pradhan, C. Narayana, and A. K. Ganguli, Nanocrystalline $NaNbO_3$ and $NaTaO_3$: Rietveld studies, Raman spectroscopy and dielectric properties, Solid State Sci. **11**, 562 (2009).

[79] S. N. Al-Refaie and S. A. Alboon, On the Analysis of $Cd_2Nb_2O_7$ Dielectric Dispersion, Phys. Procedia **25**, 15 (2012).

[80] H. Y. Zhou, X. Q. Liu, X. L. Zhu, and X. M. Chen, $CaTiO_3$ linear dielectric ceramics with greatly enhanced dielectric strength and energy storage density, Journal of the American Ceramic Society **101**, 1999 (2018).

[81] J. E. Rodrigues, D. M. Bezerra, and A. C. Hernandes, Ordering effect on the electrical properties of stoichiometric $Ba_3CaNb_2O_9$-based perovskite ceramics, Ceram. Int. **43**, 14015 (2017).